\documentclass[pre,preprint,showpacs,preprintnumbers,amsmath,amssymb,nofootinbib,showkeys]{revtex4}

\usepackage{graphicx}

\usepackage{dcolumn}

\usepackage{bm}

\begin{document}

\preprint{APS/123-QED}

\title{Energy localization on q--tori, long term stability
\vskip 5mm and the interpretation of FPU recurrences}

 \author{H. Christodoulidi$^1$}
 \email{hchrist@master.math.upatras.gr}

\author{C. Efthymiopoulos$^2$}
 \email{cefthim@academyofathens.gr}

\author{T. Bountis$^1$}
 \email{bountis@math.upatras.gr}

\affiliation{$^1$Department of Mathematics, University of Patras\\
      $^2$Research Center for Astronomy and Applied Mathematics, Academy of Athens}

\begin{abstract}

We focus on two approaches that have been proposed in recent years
for the explanation of the so-called FPU paradox, i.e. the
persistence of energy localization in the `low--$q$' Fourier modes
of Fermi--Pasta--Ulam nonlinear lattices, preventing equipartition
among all modes at low energies. In the first approach, a
low-frequency fraction of the spectrum is initially excited leading
to the formation of `natural packets' exhibiting exponential
stability, while in the second, emphasis is placed on the existence
of `$q$--breathers', i.e periodic continuations of the linear modes
of the lattice, which are exponentially localized in Fourier space.
Following ideas of the latter, we introduce in this paper the
concept of `$q$--tori' representing exponentially localized
solutions on {\it low--dimensional tori} and use their stability
properties to reconcile these two approaches and provide a more
complete explanation of the FPU paradox.

\end{abstract}


\maketitle

\section{Introduction}

In a number of recent papers, Flach and co-workers
\cite{flaetal2005, flaetal2006,flapon2008} discussed the role of
simple periodic solutions, called `$q$-breathers', in the dynamics
of the $\alpha-$ and $\beta-$versions of the Fermi-Pasta-Ulam model
with fixed boundaries. A $q$-breather is the continuation, for
$\alpha\neq 0$ or $\beta\neq 0$, of the simple harmonic motion
exhibited by the $q-$th-mode in the uncoupled case
($\alpha=\beta=0$) and implies motion by a unique frequency, which
is nearly equal to the frequency of the $q$-th mode in the uncoupled
case. For small values of the coupling parameters (and the energy),
the distribution of energy in $q$-breathers stays localized in
practice upon only a few modes. Thus, $q$-breathers offer new
insight in understanding the problem of energy localization, as well
as the long term deviations from equipartition of the energy among
the modes, i.e. the origin of the FPU paradox \cite{feretal1955}.
(For a recent and comprehensive review on discrete breathers see
\cite{flagor2008}).

Analytical estimates of various scaling laws concerning
$q$-breathers can be obtained via the method of
Poincar\'{e}-Lindstedt series \cite{flaetal2005,flaetal2006}. In particular,
expanding the solutions for all the canonical variables up to the
lowest non-trivial order with respect to the small parameters yields
an exponentially decaying function for the average harmonic energy
of the $q$th mode, $E(q)\propto\exp(-bq)$. The value of $b$ depends
on: i) $\alpha$ and/or $\beta$, ii) the number of degrees of freedom
$N$, and iii) the total energy $E$ given to the system. The most
important property of the $q$-breathers is that their energy
profiles $E(q)$ are quite similar to those of `FPU-trajectories',
i.e. solutions generated by initially exciting one or a few low-$q$
modes, whose \textit{exponential localization} has been noted since
a long time ago
\cite{galsco1972,fucetal1982,livetal1985,deletal1995}. On the basis
of this similarity it has been conjectured that there is a close
connection between $q$-breathers and the energy localization
properties of FPU-trajectories. It is intriguing, however, that this
localization persists even for values of the parameters (coupling,
energy and $N$) for which the corresponding $q$-breather solution
has become {\it unstable}. A heuristic argument for interpreting
this phenomenon was offered by Flach and Ponno \cite{flapon2008}.

Adopting a different approach, Berchialla et al. \cite{beretal2004}
explored in detail the question of energy equipartition in FPU
experiments where a constant (low-frequency) {\it fraction} of the
spectrum (instead of just one mode) was initially excited. The
numerical indication from this study, based on a limited range of
values of $E$ and $N$, was that the flow of energy across the
high-frequency parts of the spectrum takes place exponentially
slowly, by a law of the form $T\propto \exp(\varepsilon^{-1/4})$,
where $T$ is the time needed for the energy to be nearly equally
partitioned, and $\varepsilon=E/N$ is the specific energy of the
system. In fact, this dependence appears as a piecewise power law,
with different `best-fit' slopes in different ranges of values of
$\varepsilon$. For example, a power-law behavior of the type
$T\propto\varepsilon^{-3}$ was found in a subinterval of
$\varepsilon$ values considered in \cite{beretal2004}, which fits
nicely previous results on the scaling of the equipartition time
with $\varepsilon$ beyond a critical `weak chaos' threshold reported
in \cite{deletal1999}. Nevertheless, the question of whether the
scaling laws characterizing the approach to equipartition depend on
the total energy $E$, or the specific energy $E/N$ is still open,
since no rigorous results have been provided so far in the
literature. On the other hand, numerical results are available over
a limited range of values of $E$ and $N$, in which $E$ is varied
proportionally to $N$. Various semi-analytical or numerical
approaches to this question are reviewed in \cite{lichetal2008}.

Interestingly enough, even if one starts by exciting {\it one mode}
with large enough energy, one observes, long before equipartition,
the formation of \textit{metastable states} coined `natural packets'
\cite{beretal2005}, in which the energy undergoes first a kind of
equipartition among a group of low-frequency modes, as if only a
fraction of the spectrum was initially excited. We may thus conclude
that the phenomenon of metastability characterizes FPU-trajectories
resulting from all types of initial excitations of the low frequency
part of the spectrum (see \cite{benetal2008} for a review of the
`history' of the metastability scenario in the FPU problem).
Furthermore, empirical scaling laws can be established
\cite{beretal2005} concerning the dependence of the `width' of a
packet on the specific energy $\varepsilon$. These are consistent
with the laws of energy localization obtained via either a
continuous Hamiltonian model which interpolates the FPU dynamics in
the space of Fourier modes \cite{ponbam2005, bampon2006}, or the
$q$-breather model \cite{flapon2008}. A difference, however, between
the two models is that in the framework of the continuous model the
constant $b$ in $E(q)\propto\exp(-bq)$ turns out to be independent
of $E$.

The results reported in the present paper aim to provide a more
complete explanation of the FPU paradox of energy non-equipartition
by reconciling the presence of $q$-breathers and their induced
energy localization on the one hand, with the occurrence of
metastable packets of low-frequency modes on the other.

To extend the results obtained for FPU-trajectories in
\cite{deletal1999,beretal2004}, let us observe that the packet of
modes excited in these experiments corresponds to the modes with
spectral numbers satisfying the condition $(N+1)/64\leq q\leq
5(N+1)/64$, with $N$ equal to a power of two minus one. For
simplicity, let us alter this slightly and consider, instead, the
condition $1\leq q\leq 4N/64$, with $N$ a power of two. The lowest
possible value of $N$ allowed is $N=16$, for which the above
condition implies that only the $q=1$ mode is initially excited,
giving a solution close to a $q$-breather, which is an orbit lying
on an invariant {\it one-torus} of the system. Now, if $N$ is
doubled ($N=32$), the same condition implies that modes $q=1$ and
$q=2$ may now be excited, meaning that the resulting
FPU-trajectories may be regarded as lying close to an invariant {\it
two-torus} of the system $N=32$. In general, for $N=16s$, the modes
$q=1,2,\ldots,s$ are initially excited and the motion should be
regarded as lying close to an invariant $s$-torus of the system. The
same holds true when natural packets of width $s$ are formed in
experiments in which the initial conditions are as adopted in
\cite{beretal2005}.

This leads to the idea that the properties of the FPU-trajectories
could be understood by considering {\it classes} of special
solutions lying not only on one-dimensional tori (as is the case
with $q$-breathers), but also on tori of {\it any low dimension}
$s<<N$, i.e. solutions with $s$ independent frequencies,
representing the continuation of motions resulting from exciting $s$
modes of the uncoupled case. Generalizing the concept of
$q$-breathers, we call {\it $q$-tori} the quasiperiodic solutions
solutions on such low-dimensional tori. The main body of the present
paper, therefore, focuses on exploring the properties of these
$q$-tori solutions, both analytically and numerically. In
particular, we establish for $q$-tori energy localization laws
analogous to those for $q$-breathers, using a semi-analytical
approach. Our numerical experiments then show that such laws
describe accurately the properties not only of exact $q$-tori
solutions, but also of FPU trajectories with nearby initial
conditions.

Our work shares a common starting point with a recent paper by
Giorgilli and Muraro \cite{giomur2006}, where the authors have also
explored the idea of FPU trajectories being confined on
lower-dimensional manifolds embedded in the $2N$-dimensional FPU
phase space. Their main result, proving that the confinement
persists for times exponentially long in the inverse of $\epsilon$,
is obtained in the spirit of the theory of Nekhoroshev
\cite{nek1977, benetal1985}, using a variant of the formulation of
the Nekhoroshev  theorem for `isochronous' systems
\cite{gio1988,fasetal1998,guzetal1998, nie1998,eftetal2004}. In
fact, the theory of Nekhoroshev appears to offer quite a natural
framework for studying analytically the metastability scenario.

However, as has been pointed out quite early \cite{gio1992} a naive
application of the Nekhoroshev theory in the FPU problem would break
down as $N\rightarrow\infty$, since, under the assumption that the
Nekhoroshev time $T$ depends on the specific energy $\varepsilon$,
in the estimates $T\sim\exp(1/\varepsilon^c)$ of the theory the
exponent is of the form $c=O(1/N)$ (see \cite{eftetal2004} for a
heuristic explanation). This bad dependence of $c$ on $N$ actually
implies that $T=O(1)$ as $N\rightarrow\infty$, hence Nekhoroshev's
theory fails to predict long times for equipartition in that limit
(a relevant theoretical result has only been obtained in lattices
where a clear separation of the frequencies occurs into a low and
high band; the low frequencies then become small parameters, see
\cite{benetal1987}). In \cite{giomur2006} one has instead
$c=O(1/m)$, where $m$ is one-half the dimension of the
lower-dimensional manifold where confined FPU trajectories are
expected to lie. This constitutes a significant improvement with
respect to previous estimates, but still does not explain the
natural packets correctly, since the latter's width was found in
numerical experiments to vary as $m\propto \varepsilon^{1/4} N$
\cite{beretal2005}, i.e. proportionally to $N$, for fixed
$\varepsilon$.

Our analytical theory relies on the use of the {\it Poincar\'{e} -
Lindstedt} method, through which we find scaling laws for the energy
profile $E(q)$ of a trajectory lying exactly on a $q$-torus. The
consistency of the Poincar\'{e}-Lindstedt construction on a Cantor
set of perturbed frequencies (or amplitudes) is explicitly
demonstrated. Numerically, we find that energy localization persists
for appreciably long times, for trajectories neighboring a
$q$-torus, even beyond the energy threshold where the $q$-torus
becomes linearly unstable. The determination of linear stability for
an $s$-dimensional torus ($s>1$) is, of course, a subtle question,
since no straightforward application of Floquet theory is available,
as in the $s=1$ case. Nevertheless, by employing an effective and
reliable criterion for the stability of $q$-tori via the use of the
recently developed method of `Generalized Alignment Indices' (GALI,
see \cite{skoetal2007}), we are able to determine approximate
critical parameter values at which a low-dimensional torus turns
unstable, in the sense that orbits in its vicinity display chaotic
behavior.

It is important to remark at this point that, although the
Poincar\'{e}-Lindstedt approach is quite distinct from the Birkhoff
method used in Nekhoroshev theory, it appears that the properties of
exponential localization can be exploited to demonstrate (by
subsequent use of the Birkhoff method), an even better behavior of
the exponent $c$ of Nekhoroshev estimates than can be found in the
literature so far. A detailed exploration of this issue is deferred
to a separate study. Nevertheless, a heuristic argument offered in
the closing section of this paper suggests that the removal of the
dependence of $c$ on the number of degrees of freedom is possible,
at least for trajectories close to $q$-breathers.

Our paper is structured as follows: Section 2 presents our
analytical results on the existence and scaling laws of the
`$q$-tori'. We deal here only with the $\beta-$case, but our
approach can be readily extended to the $\alpha-$case as well. We
focus on the specific and most relevant subset of `$q$-tori'
corresponding to zeroth order excitations of a set of adjacent modes
$q_{0,i}=i$, $i=1,\ldots,s$, whose energy profile $E(q)$ is
calculated analytically. Next, our analytical predictions are tested
against numerical integration of specific orbits.  Section 3
examines the question of stability of $q$-tori and the persistence
of energy localization of the FPU-trajectories when the linear
stability of the `underlying' $q$-tori is lost. Finally, we deal
with the question of the long-term stability of exponentially
localized FPU-trajectories, via heuristic estimates inspired by
Nekhoroshev's theory. Section 4 summarizes the main conclusions of
the present study.

\section{Existence and stability of $q$-tori}

\subsection{The FPU $\beta$--model}
The $\beta-$FPU Hamiltonian for a lattice of $N$ particles reads:
\begin{equation}\label{fpuham}
H= {1\over 2}\sum_{k=1}^N y_k^2 + {  1   \over
2}\sum_{k=0}^N(x_{k+1}-x_k)^2 + {\beta \over 4}
\sum_{k=0}^N(x_{k+1}-x_k)^4
\end{equation}
where $x_k$ is the $k$-th particle's displacement with respect to
the equilibrium position and $y_k$ is its canonically conjugate
momentum. Fixed boundary conditions are defined by setting
$x_0=x_{N+1}=0$.

The normal mode canonical variables $(Q_q,P_q)$ are introduced by
the linear canonical transformations
\begin{equation}\label{lintra}
x_k=\sqrt{2\over N+1}\sum_{q=1}^N Q_q\sin\left({qk\pi\over
N+1}\right),~~~ y_k=\sqrt{2\over N+1}\sum_{q=1}^N
P_q\sin\left({qk\pi\over N+1}\right)
\end{equation}
Substitution of (\ref{lintra}) into (\ref{fpuham}) yields the
Hamiltonian in the form $H=H_2+H_4$ in which the quadratic part is
diagonal
\begin{equation}\label{fpuham2}
H_2=\sum_{q=1}^N {P_q^2+\Omega_q^2Q_q^2\over 2}
\end{equation}
with normal mode frequencies
\begin{equation}\label{fpuspec}
\Omega_q=2\sin\left({q\pi\over 2(N+1)}\right),~~~1\leq q\leq
N~~~.
\end{equation}
The quartic part of the Hamiltonian reads
\begin{equation}\label{fpuham4}
H_4={\beta\over 2(N+1)}\sum_{q,l,m,n=1}^N C_{q,l,m,n}
\Omega_q\Omega_l\Omega_m\Omega_n
Q_qQ_lQ_mQ_n
\end{equation}
where the coefficients $C_{q,l,m,n}$ take non-zero values only for
particular combinations of the indices $q,l,m,n$, namely
\begin{equation}\label{ccoef}
C_{q,l,m,n}= \left\{
\begin{array}{rl}
 1 & \mbox{if}~q \pm l \pm m \pm n = 0\\
-1 & \mbox{if}~q \pm l \pm m \pm n = \pm 2(N+1)
\end{array}
\right.
\end{equation}
in which all possible combinations of the $\pm$ signs are taken
into account. Thus, in the new canonical variables, the equations
of motion are:
\begin{equation}\label{eqmo}
\ddot{Q}_q+\Omega_q^2Q_q=-{\beta\over
2(N+1)}\sum_{l,m,n=1}^N C_{q,l,m,n}
\Omega_q\Omega_l\Omega_m\Omega_n
Q_lQ_mQ_n~~~.
\end{equation}

If $\beta=0$, the individual harmonic energies
$E_q=(P_q^2+\Omega_q^2Q_q^2)/2$ are preserved by Eqs.(\ref{eqmo}),
i.e. the energies $E_q$ form a set of $N$ integrals in involution.
When $\beta\neq 0$, however, the harmonic energies become
functions of time and only the total energy $E=\sum_{q=1}^N
E_q(t)$ is conserved. The specific energy is then defined as
$\varepsilon=E/N$, while the average harmonic energy of each mode
over a time interval $0\leq t\leq T$ is given by the integral
$\bar{E}_q(T)={1\over T}\int_0^T E_q(t) dt$.

In classical FPU experiments, one starts with the total energy
shared only by a small subset of modes. Then, for short time
intervals $T$, we have $E_q(T)\simeq 0$ for all $q$ corresponding
to non-excited modes. Equipartition means that, due to the
nonlinear terms, the energy will eventually be shared equally by
all modes, i.e.
\begin{equation}\label{equip}
\lim_{T\rightarrow\infty}\bar{E}_q(T) =
\varepsilon~~,\,\,q=1,...,N.
\end{equation}
The usual ergodic assumption of statistical mechanics leads to the 
conclusion that (\ref{equip}) is violated only for orbits resulting 
from a zero measure set of initial conditions. The FPU paradox owes 
its name to the crucial observation that large deviations from the
approximate equality $\bar{E}_q(T)\simeq\varepsilon$ occur for
many other orbits as well. Depending on the initial conditions,
these deviations are termed `FPU recurrences' and are seen to
persist even when $T$ becomes very large.

\subsection{$q$-tori and their construction by Poincar\'{e}-Lindstedt series}

From the above discussion, one infers that the dynamics of FPU
recurrences is governed by particular solutions of the FPU equations
(\ref{eqmo}) for which energy localization occurs only on a small
subset of Fourier modes. Such solutions lie on {\it tori of
low--dimensionality}, which we shall henceforth call `$q$--tori',
since they also turn out to be exponentially localized in Fourier
space, like the `$q$--breather' solutions discussed in
\cite{flaetal2005, flaetal2006}.

We now introduce the main ingredients of our method of $q$-torus
construction, using an explicitly solved example for $N=8$, whose
solutions lie on a two-dimensional torus representing the
continuation, for $\beta\neq 0$, of the quasiperiodic solution of
the uncoupled ($\beta=0$) system $Q_1(t)=A_1\cos\Omega_1 t$,
$Q_2=A_2\cos\Omega_2 t$, for a suitable choice of $A_1$ and $A_2$.

To this end, we follow the Poincar\'{e} - Lindstedt method and look
for solutions $Q_q(t)$, $q=1,\ldots,8$ expanded as series in the
parameter $\sigma=\beta/2(N+1)$, namely
\begin{equation}\label{qser}
Q_q(t)=Q_q^{(0)}(t)+\sigma Q_q^{(1)}(t) + \sigma^2
Q_q^{(2)}(t)+\ldots,~~~q=1,\ldots 8~~~.
\end{equation}
For the motion to be quasiperiodic on a two--torus, the functions
$Q_q^{(r)}(t)$ must, at any order $r$, be trigonometric polynomials
involving only two frequencies (and their multiples). Furthermore,
for the motion to represent a continuation of the unperturbed
solutions $Q_1$ and $Q_2$, the frequencies $\omega_1$ and $\omega_2$
must be small corrections of the normal mode frequencies $\Omega_1$,
$\Omega_2$. According to the Poincar\'{e} - Lindstedt method, these
new frequencies are also given by series in powers of $\sigma$, as:
\begin{equation}\label{omeser}
\omega_q=\Omega_q +
\sigma\omega_q^{(1)}+\sigma^2\omega_q^{(2)}+\ldots~~~q=1,2~~.
\end{equation}
The corrections are determined by the requirement that all terms in
the differential equations of motion, giving rise to secular terms
(of the form $t\sin\omega_qt$ etc.) in the solutions $Q_q(t)$, be
eliminated.

Let us consider the equation of motion for the first mode, whose
first few terms on the right hand side (r.h.s) are:
\begin{equation}\label{eqq1}
\ddot{Q}_1+\Omega_1^2Q_1=-\sigma(3\Omega_1^4Q_1^3
+6\Omega_1^2\Omega_2^2Q_1Q_2^2
+3\Omega_1^3\Omega_3Q_1^2Q_3+...)
\end{equation}
Proceeding with the Poincar\'{e} - Lindstedt series, the frequency
$\Omega_1$ is substituted on the l.h.s. of Eq.(\ref{eqq1}) by its
equivalent expression obtained by squaring (\ref{omeser}) and
solving for $\Omega_1^2$. Up to first order in $\sigma$ this gives
\begin{equation}\label{ome1inv}
\Omega_1^2=\omega_1^2-2\sigma\Omega_1\omega_1^{(1)}+...
\end{equation}
Substituting the expansions (\ref{qser}) into (\ref{eqq1}),
as well as the frequency expansion (\ref{ome1inv}) into the l.h.s. of
(\ref{eqq1}), and grouping together terms of like orders, we find at
zeroth order $\ddot{Q}_1^{(0)}+\omega_1^2Q_1^{(0)}=0$, while at first
order
\begin{eqnarray}\label{eqq11}
\ddot{Q}_1^{(1)}+\omega_1^2Q_1^{(1)}&=&2\Omega_1\omega_1^{(1)}Q_1^{(0)}
-3\Omega_1^4(Q_1^{(0)})^3-
6\Omega_1^2\Omega_2^2Q_1^{(0)}(Q_2^{(0)})^2\nonumber\\
&-&3\Omega_1^3\Omega_3(Q_1^{(0)})^2Q_3^{(0)}+...
\end{eqnarray}
Repeating the above procedure for modes 2 and 3, we find that their
zeroth order equations also take the harmonic oscillator form:
\begin{equation}\label{eqq2030}
\ddot{Q}_2^{(0)}+\omega_2^2Q_2^{(0)}=0,~~~~~~
\ddot{Q}_3^{(0)}+\Omega_3^2Q_3^{(0)}=0~~~~.
\end{equation}
Note that the {\it corrected} frequencies $\omega_1$, $\omega_2$ appear in the
zeroth order equations for the modes 1 and 2, while the {\it uncorrected} frequency
$\Omega_3$ appears in the zeroth order equation of the mode 3 (similarly,
$\Omega_4,\ldots,\Omega_8$ appear in the zeroth order equations of the modes
4 to 8). Continuing the construction of a solution which lies on a two-torus,
we start from particular solutions of (\ref{eqq2030}) (with zero velocities
at $t=0$) which read:
$$
Q_1^{(0)}(t)=A_1\cos\omega_1t,~~ Q_2^{(0)}(t)=A_2\cos\omega_2t,~~
Q_3^{(0)}(t)=A_3\cos\Omega_3t
$$
where the amplitudes $A_1,A_2,A_3$ are arbitrary. If the solution
is to lie on a two-torus with frequencies $\omega_1$, $\omega_2$,
we must set $A_3=0$, so that no third frequency is introduced in
the solutions. In the same way, the zeroth order equations
$\ddot{Q}_q^{(0)}+\Omega_q^2Q_q^{(0)}=0$ for the remaining modes
$q=4,...8$ yield solutions $Q_q^{(0)}(t)=A_q\cos\Omega_qt$ and we
set $A_4=A_5=...=A_8=0$. Thus, we are left with only two non-zero
free amplitudes $A_1$, $A_2$.

Now, consider Eq.(\ref{eqq11}) for the first order term
$Q_1^{(1)}(t)$. Only zeroth order terms $Q_q^{(0)}(t)$ appear on
its r.h.s., allowing for the solution to be found recursively. The
crucial remark is that by the choice $A_3=\ldots=A_8=0$, one also
has $Q_3^{(0)}(t)=\ldots=Q_8^{(0)}(t)=0$, whence only a small
subset of the terms appearing in the original equations of motion
survive on the r.h.s. of Eq.(\ref{eqq11}), namely those in which
none of the functions $Q_3^{(0)}(t), \ldots, Q_8^{(0)}(t)$
appears. As a result, equation (\ref{eqq11}) is simplified
dramatically and upon substitution of
$Q_1^{(0)}(t)=A_1\cos\omega_1t$, $Q_2^{(0)}(t)=A_2\cos\omega_2t$
reduces to:
\begin{eqnarray}\label{eqq11n}
\ddot{Q}_1^{(1)}+\omega_1^2Q_1^{(1)}&=&2\Omega_1\omega_1^{(1)}A_1\cos\omega_1t
-3\Omega_1^4A_1^3\cos^3\omega_1t\nonumber\\
&-&6\Omega_1^2\Omega_2^2A_1A_2^2\cos\omega_1t\cos^2\omega_2t~~~.
\end{eqnarray}
This can now be used to fix $\omega_1^{(1)}$ so that no secular
terms appear in the solution, yielding
$$
\omega_1^{(1)}={9\over 8}A_1^2\Omega_1^3 +{3\over
2}A_2^2\Omega_1\Omega_2^2
$$
while, after some simple operations, we find for $Q_1^{(1)}$:
\begin{eqnarray}\label{q11sol}
Q_1^{(1)}(t)&=&{3A_1^3\Omega_1^4\cos 3\omega_1t\over
32\omega_1^2}+{3A_1A_2^2\Omega_1^2\Omega_2^2\cos
(\omega_1+2\omega_2)t\over
2[(\omega_1+2\omega_2)^2-\omega_1^2]}\nonumber\\
&+&{3A_1A_2^2\Omega_1^2\Omega_2^2\cos
(\omega_1-2\omega_2)t\over 2[(\omega_1-2\omega_2)^2-\omega_1^2]}~~.
\end{eqnarray}
By the same analysis, we fix the frequency correction of the
second mode:
$$
\omega_2^{(1)}={9\over 8}A_2^2\Omega_2^3 +{3\over
2}A_1^2\Omega_1^2\Omega_2
$$
and obtain the solution
\begin{eqnarray}\label{q21sol}
Q_2^{(1)}(t)&=&{3A_2^3\Omega_2^4\cos 3\omega_2t\over
32\omega_2^2}+{3A_1^2A_2\Omega_1^2\Omega_2^2\cos
(2\omega_1+\omega_2)t\over
2[(2\omega_1+\omega_2)^2-\omega_2^2]}\nonumber\\
&+&{3A_1^2A_2\Omega_1^2\Omega_2^2\cos
(2\omega_1-\omega_2)t\over 2[(2\omega_1-\omega_2)^2-\omega_2^2]}~~
\end{eqnarray}
which has a similar structure as the first order solution of the
first mode. For the third order term there is no frequency
correction, and we find
\begin{eqnarray}\label{q31sol}
&~&Q_3^{(1)}(t)= {A_1^3\Omega_1^3\Omega_3\over
4}\left({3\cos \omega_1t\over\omega_1^2-\Omega_3^2} + {\cos
3\omega_1t\over
9\omega_1^2-\Omega_3^2}\right) \nonumber\\
&+&
{3A_1A_2^2\Omega_1\Omega_2^2\Omega_3\over
4}\left({\cos (\omega_1-2\omega_2)t\over
(\omega_1-2\omega_2)^2-\Omega_3^2} + {\cos
(\omega_1+2\omega_2)t\over (\omega_1+2\omega_2)^2-\Omega_3^2}+{2\cos
\omega_1t\over \omega_1^2-\Omega_3^2}\right)~~.
\end{eqnarray}
We may thus proceed to the sixth mode to find solutions in which all
the functions $Q_3^{(0)},\ldots,Q_6^{(0)}$ are equal to zero, while
the functions $Q_3^{(1)},\ldots,Q_6^{(1)}$ are non zero. However, a
new situation appears when we arrive at the seventh and eighth
modes. A careful inspection of the equation for the term
$Q_7^{(1)}(t)$
\begin{equation}\label{eqq71}
\ddot{Q}_7^{(1)}+\Omega_7^2Q_7^{(1)}=-\sum_{l,m,n=1}^8
C_{7,l,m,n} \Omega_7\Omega_l\Omega_m\Omega_n
Q_l^{(0)}Q_m^{(0)}Q_n^{(0)}~~~
\end{equation}
shows that there can be {\it no term} on the r.h.s. which does not
involve some of the functions $Q_3^{(0)},\ldots,Q_8^{(0)}$. Again,
this follows from the selection rules for the coefficients of
Eq.(\ref{ccoef}). Since all these functions are equal to zero,
the r.h.s. of Eq.(\ref{eqq71}) is equal to zero. Taking this into
account, we set $Q_7^{(1)}(t)=0$, so as not to introduce a third
frequency $\Omega_7$ in the solutions, whence the series expansion
(\ref{qser}) for $Q_7(t)$ necessarily starts with terms of order at
least  $O(\sigma^2)$. The same holds true for the equation determining
$Q_8^{(1)}(t)$.

Some remarks regarding the above construction are in order:

i) {\it Consistency}. The solutions (\ref{q11sol}) -- (\ref{q31sol})
(and those of subsequent orders) are meaningful only when the
frequencies appearing in the denominators satisfy no
commensurability condition. The spectrum of uncorrected frequencies
$\Omega_q$, given by (\ref{fpuspec}), is fully incommensurable only
if $N$ is either a prime number minus one, or a power of two minus
one \cite{hem1959}. In all other cases, there are commensurabilities
among the unperturbed frequencies, examples of which are given e.g.
in \cite{rink2001}. However, such commensurabilities do not affect
the consistency of the construction of the Poincar\'{e}-Lindstedt
series, because it can be shown that no divisors of the form
$\sum_{q=1}^N n_q\Omega_q$ appear in the series at any order and for
any integer vector
$\mathbf{n}\equiv(n_1,n_2,\ldots,n_q)\neq(0,0,\ldots,0)$. The proof
of this statement follows from the fact that the kernel differential
equations (like Eq.(\ref{eqq11n})) determining all terms
$Q_q^{(k)}$, $q=1,\ldots 8$ at order $k$ read:
\begin{eqnarray}\label{kernpl}
\ddot{Q}_q^{(k)}(t)&+&\omega_q^2Q_q^{(k)}(t)=\nonumber\\
& &B_{q}^{(k)}\omega_q^{(k)}\cos\omega_q t
+\sum_{\mathop{n_1,n_2\in Z}\limits_{|n_1|+|n_2|\neq 0}}^{k-1}
A_{q,n_1,n_2}^{(k)}\cos[(n_1\omega_1+n_2\omega_2)t]~~~\mbox{if}~q=1,2
\label{QkEq}
\end{eqnarray}
while for $q=3,...,8$ the same equation holds with $B_{q}^{(k)}=0$
and  $\omega_q^2$ on
the l.h.s. replaced by $\Omega_q^2$. The coefficients $B_q^{(k)}$
and $A_{q,n_1,n_2}^{(k)}$ in (\ref{QkEq}) are determined at
previous steps of the construction (see Eq.(\ref{mod}) below for
arbitrary $s$). It follows, therefore, that all new divisors
appearing at successive orders, belong to one of the following
sets:
$$
(n_1\pm 1)\omega_1+n_2\omega_2,~~~
n_1\omega_1+(n_2\pm 1)\omega_2,~~~
\Omega_q\pm(n_1\omega_1+n_2\omega_2),~q=3,\ldots 8
$$
for $n_1,n_2\in\mathbb{Z}$, $|n_1|+|n_2|\neq 0$, whence we deduce
that no zero divisors can ever show up in the Poincar\'{e}-Lindstedt
series due to commensurabilities between the unperturbed frequencies
$\Omega_q$. On the other hand, we can also exclude the appearance of
zero divisors due to the perturbed frequencies $\omega_1,\omega_2$,
if the frequencies in a quasiperiodic solution are {\it fixed in
advance} so that the two frequencies do not belong to the countable
set of all planes defined by the relations $(n_1\pm
1)\omega_1+n_2\omega_2=0$, $n_1\omega_1+(n_2\pm 1)\omega_2=0$,
$\Omega_q+(n_1\omega_1+n_2\omega_2)=0$, $q=3,\ldots 8$, for all
integer values $n_1,n_2\in Z$, $|n_1|+|n_2|\neq 0$. We stress that
fixing the frequencies in advance is a necessary ingredient of the
Poincar\'{e} - Lindstedt method, since otherwise the equations at all
orders (as, for example, Eqs.(\ref{q11sol}), (\ref{q21sol}) and
(\ref{q31sol})) would not be solvable. Furthermore, this frequency
specification is analogous to the procedure followed in the
construction of Kolmogorov's normal form representing solutions on
KAM tori (see \cite{gio1998} for a detailed comparison of the two
methods). Since the frequencies $\omega_1$, $\omega_2$ are functions
of the amplitudes $A_1,A_2$, one deduces that the formal consistency
of the method can be established in the complement of all excluded
planes, i.e. a Cantor set in either the frequency space
$(\omega_1,\omega_2)$ or the amplitude space $(A_1,A_2)$.

The above demonstration of consistency is readily generalized to the
construction of $s$-dimensional $q$-tori by the nonlinear
continuation of the set of linear modes $q_i$, $i=1,...,s$. Namely,
one can demonstrate that the consistency holds on a Cantor set of
amplitude multiplets $(A_{q_1},A_{q_2},\ldots,A_{q_s})$ such that
the resulting perturbed frequencies $\omega_{q_i}, i=1,\ldots,s$ not
lie on any one of the planes:
$$
(n_1+m_1)\omega_{q_1}+(n_2+ m_2)\omega_{q_2}+\ldots+(n_s+m_s)\omega_{q_s}=0
$$
or
$$
\Omega_q\pm(n_1\omega_{q_1}+n_2\omega_{q_2}+\ldots+n_s\omega_{q_s})=0
$$
where $q=\{1,\ldots,N\}\backslash\{q_1,\ldots,q_s\}$,
$n_1,\ldots,n_s\in N$, $|n_1|+\ldots+|n_s|\neq 0$ and $m_q=-1$, $0$
or $1$ for all $q=1,\ldots, s$. For example, if the first $s$ modes
are excited by amplitudes $A_k$, $k=1,...s$, the formula for the
perturbed frequencies reads
\begin{eqnarray}\label{freqs}
 \omega_{q}&=&\Omega_q+\frac{3\sigma}{2} \Omega_{q}
\sum_{k=1}^{s}A_{k}^{2}\Omega_{k}^{2} -\frac{3}{8}
A_{q}^{2}\Omega_{q}^{3}+O(\sigma^2;A_1,...A_s),~~~q=1,\ldots,s
\end{eqnarray}
Fixing the values of the frequencies $\omega_q$ in advance implies
that Eqs.(\ref{freqs}) should be regarded as yielding the (unknown)
amplitudes $A_k$ for which the quasiperiodic solution exhibits the
chosen set of frequencies on a Cantor set in the space of the
amplitudes $A_1,\ldots,A_s$. The case $s=1$ implies that the same
property holds for the Poincar\'{e} - Lindstedt series representing
$q$-breathers \cite{flaetal2006}. That is, while by Lyapunov's
theorem the continuation of the periodic orbits is guaranteed in an
open domain of values $A_{q0}$, Poincar\'{e}-Lindstedt series can
only be constructed on a Cantor subset of this domain. Yet, this is
sufficient for our present purpose, which is to determine by a
semi-analytical approach scaling laws for the energy localization on
$q$-tori (or $q$-breathers).

ii) {\it Convergence}. Even after consistency is demonstrated, no
proof has yet been provided for the convergence of the series. As
demonstrated in the works of Eliasson \cite{eli1997} and Gallavotti
\cite{gal1994a,gal1994b}, the question of convergence of the
Lindstedt series is notoriously difficult even in simple Hamiltonian
systems. This is because the Lindstedt series for orbits on invariant
tori are convergent, but not absolutely (see the review by Giorgilli
\cite{gio1999}). On the other hand, it is possible to make an
absolutely convergent classical expansion by use of the Kolmogorov
normal form as developed by Giorgilli and Locatelli
\cite{gioloc1997}. Such analysis, however, is quite cumbersome and
will be deferred to another publication as it would obscure the
results presented here.

\begin{figure}
\centering
\includegraphics[scale=0.55]{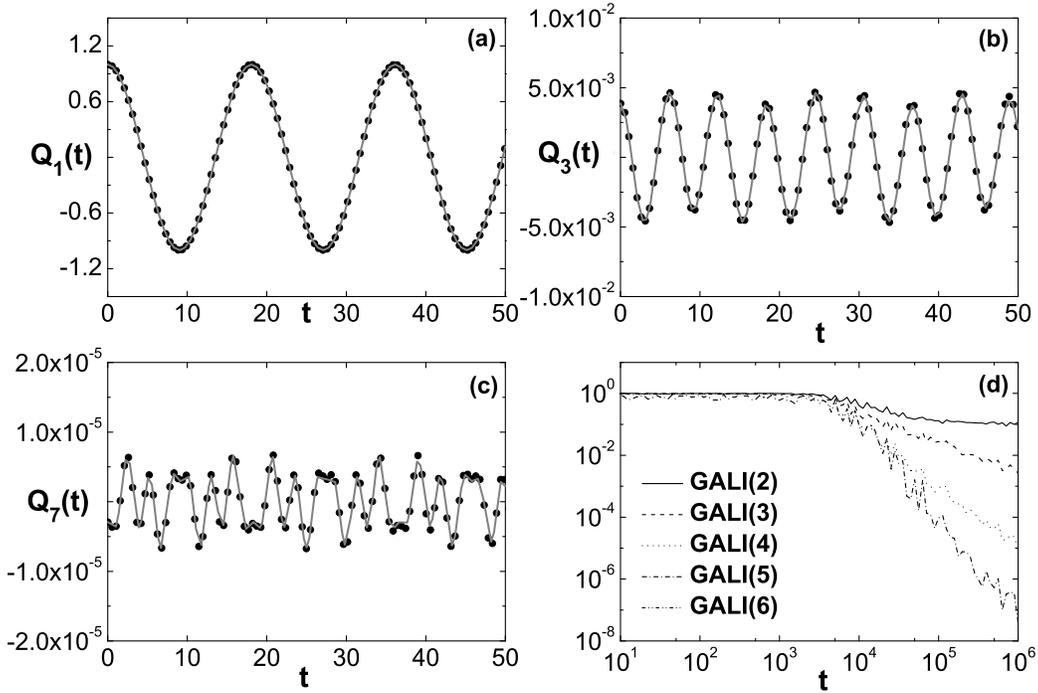}
\caption{Comparison of numerical (points) versus analytical (solid
line) solutions, using the Poincar\'{e} - Lindstedt series up to
order $O(\sigma^2)$, for the temporal evolution of the modes (a)
$q=1$, (b) $q=3$, (c) $q=7$, when $A_1=1$, $A_2=0.5$, and $N=8$,
$\beta=0.1$ and (d) the time evolution of the indices $GALI_2$ to
$GALI_6$ up to a time $t=10^6$ show that the motion lies an
2--dimensional torus.} \label{qtor8}
\end{figure}
Thus, we prefer to justify our previous statements by numerical
simulations, taking our initial conditions from the analytical
solutions (\ref{qser}) at $t=0$ and using the GALI method
\cite{skoetal2007} to demonstrate that the solutions lie, indeed, on
two-dimensional tori, as shown in Figure \ref{qtor8}. The numerical
solution for the modes $Q_1(t),Q_3(t)$ and $Q_7(t)$ is checked
against the analytical solution via the Poincar\'{e}-Lindstedt
series, truncated at second order with respect to $\sigma$, when
$N=8$, $\beta=0.1$, and $A_1=1$, $A_2=0.5$. The size of the error is
found to be precisely that expected by the truncation order in
Fig.\ref{qtor8}a-c, while in Fig.\ref{qtor8}d the GALI method shows
that the numerical orbit lies on a 2-torus.

Let us recall that according to Skokos et al. \cite{skoetal2007},
the indicators $GALI_k$, $k=2,3,...$, for chaotic orbits, decay
exponentially fast, due to the attraction of all deviation vectors
by the most unstable direction corresponding to the maximal Lyapunov
exponent. On the other hand, if an orbit lies on a {\it stable}
s-dimensional torus, the indices $GALI_2,\ldots,GALI_s$ oscillate
about a {\it non-zero} value, while the $GALI_{s+j}, j=1,2,..$
follow asymptotically power laws falling at least as $t^{-j}$.

This is precisely what we observe in Fig. \ref{qtor8}d. Namely,
after a transient initial interval (required for phase mixing to
become effective), the $GALI_2$ index stabilizes at a constant value
$GALI_2\simeq 0.1$, while all subsequent indices, starting from
$GALI_3$ decay following a power law as predicted by the theory.
The time for phase mixing is estimated to be of order
$N^3/\beta\approx 10^4$. This implies that beyond $t=10^4$ we should
observe the expected asymptotic behavior of the GALI indices to set
in, as is the case in Fig.\ref{qtor8}d. Thus, we conclude that the
motion lies on a 2-torus, exactly as predicted by the
Poincar\'{e}-Lindstedt construction, despite the fact that some
excitation was provided initially to all modes.

(iii) {\it Presence and accumulation of small divisors}. There are,
of course, small divisors appearing in terms of all orders beyond
the zeroth. First, the low-mode frequencies satisfy $\omega_q\sim
\pi q/N$, and hence divisors like $\omega_1^2$, or
$(\omega_1-2\omega_2)^2 \pm\omega_1^2$ (appearing e.g. in
Eq.(\ref{q11sol})) are small and care must be taken regarding their
effect on the growth of terms of the series at successive orders. In
fact, the most important effects are introduced by {\it nearly
resonant} divisors, like $9\omega_1^2-\Omega_3^2$ in
Eq.(\ref{q31sol}). Since the first order corrections of the
frequencies $\omega_1$ and $\omega_2$ are $O(\beta A_j^2/N^4)$, for
$A_j$, $\beta$ sufficiently small, one may still use for them the
approximation given by the first two terms in the sinus expansion of
(\ref{fpuspec}) namely:
\begin{equation}\label{fpures}
\omega_q \simeq {\pi q\over (N+1)}-{\pi^3 q^3\over 24(N+1)^3}
\end{equation}
the error being $O(A_j^2\beta/N^4)$ for $\omega_1,\omega_2$, and
$O((q/N)^5)$ for all other frequencies. This implies that a divisor
like $9\omega_1^2-\Omega_3^2$ can be approximated by the relation
$$
|q^2\omega_1^2-\Omega_q^2|=|(q\omega_1+\Omega_q)(q\omega_1-\Omega_q)|\simeq
{2q\pi\over (N+1)}{\pi^3(q^3-q)\over
24(N+1)^3}\simeq {\pi^4q^4\over 12N^4}+O\left({q^6\over N^6}\right)~~.
$$
In general, the terms produced at consecutive orders involve
products of divisors, whose influence on the size of the series
terms must be taken into account as regards estimates of the profile
of energy localization for the $q$-tori solutions, as explained in
subsection II.D below.

\begin{figure}
\centering
\includegraphics[scale=0.55]{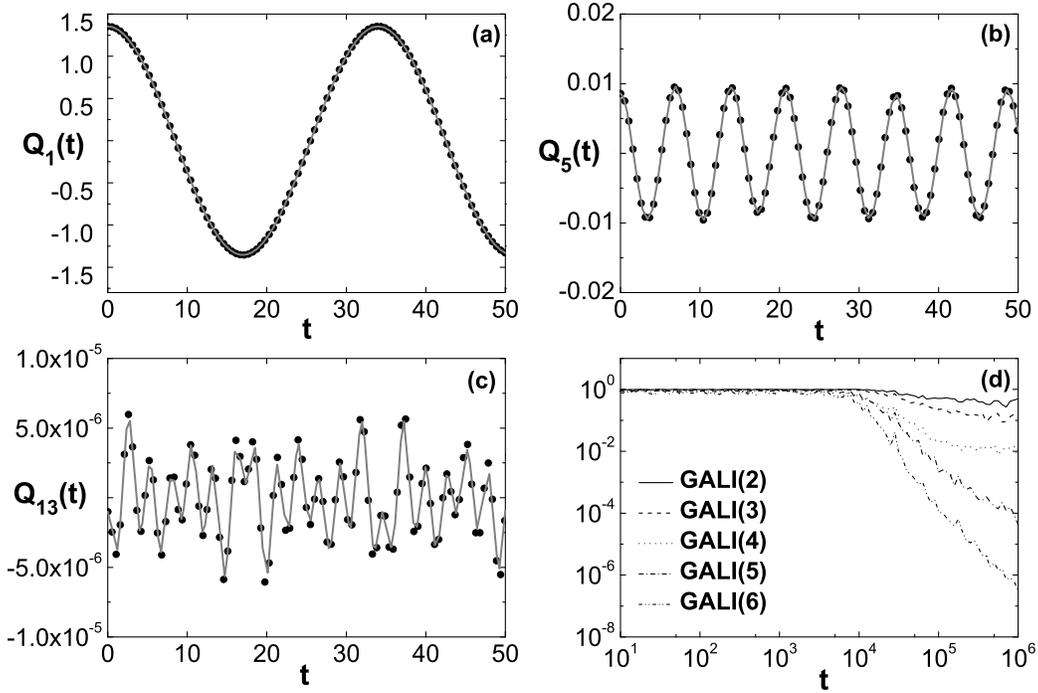}
\caption{Same as in Fig. \ref{qtor8}, showing the existence of a
4-torus in the system with $N=16$, $\beta=0.1$, when $A_1=1$,
$A_2=0.5$, $A_3=0.333...$, $A_4 = 0.25$. The temporal evolution
$Q_q(t)$ is shown for the modes (a)$q=1$, (b)$q=5$, (c)$q=13$. (d)
The $GALI_k$, $k=2,3,4$ indices, are seen to stabilize after $t\sim
10^4$, while for $k\geq 5$ the indices continue to decrease by power
laws.} \label{qtor16}
\end{figure}
(iv) {\it Sequence of mode excitations}. The profile of energy
localization along a $q$-torus solution is determined by the
sequence of mode excitations arising as the recursive scheme
proceeds to subsequent orders. The term `excitation' here means that
the solutions of the Poincar\'{e} - Lindstedt method should be
non-zero for the first time at the order where it is claimed that
the excitation takes place. For example, as already explained in our
previous construction of a 2-torus solution, the modes 1,2 are
excited at zeroth order, the modes 3, 4, 5, and 6 at first order,
and the modes 7 and 8 at the second order of the recursive scheme.

Figure \ref{qtor16} presents a one more example showing the
comparison between our analytical and numerical results for the
modes $Q_1(t)$, $Q_5(t)$, and $Q_{13}(t)$, along a 4-torus solution
constructed precisely as described above, with $N=16$, $\beta=0.1$
and by exciting modes 1 -- 4 at zeroth order, via the amplitudes
$A_1=1$, $A_2=0.5$, $A_3=0.333\ldots$, $A_4=0.25$. In this case, we
find that the modes excited at the first order of the recursive
scheme are $q=5$ to $q=12$, while the modes excited at second order
are $q=13$ to $q=16$ and the GALI method shows that the motion
occurs on a 4-torus, since $GALI_5$ is the first index to drop
asymptotically like $t^{-1}$ (see Fig. \ref{qtor16}d).

The sequence of excitation of different modes plays demonstrably a
crucial role in estimating the profile of energy localization, since
the amplitudes of all excited modes at the $r$-th order have a
pre-factor $\sigma^r=(\beta/2(N+1))^r$. This is the subject of the
next subsection, in which a proposition is provided for the sequence
of mode excitations, in the generic case of {\it arbitrary} $N$ and
{\it arbitrary} dimension $s$ of the low-dimensional torus (with the
restriction $s<N$). Next, the consequences of this proposition are
examined on the localization profile of $q$-tori and nearby FPU
trajectories.

\subsection{Sequence of mode excitations}

\begin{figure}
\centering
\includegraphics[scale=0.55]{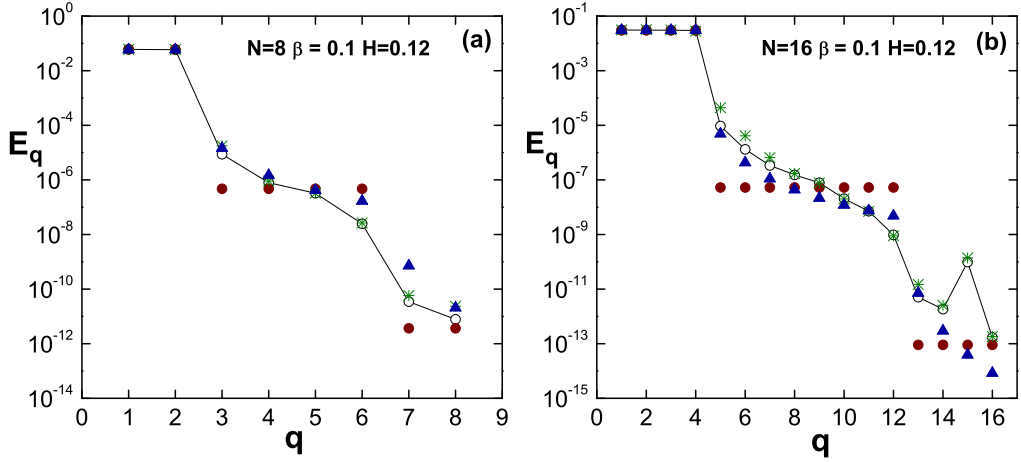}
\caption{(Color
 online)  The average harmonic energy $E_q$ of the $q$-th mode as a
function of $q$, after a time $T=10^6$ for (a) the 2-torus and (b)
4-torus solutions (open circles) corresponding to the initial
conditions used in Figs.\ref{qtor8} and \ref{qtor16} respectively.
The stars are $E_q$ values calculated via the analytical
representation of the solutions $Q_q(t)$ by the Poincar\'{e}
-Lindstedt series. The filled circles show a theoretical estimate
based on the average energy of suitably defined groups of modes (see
Eq.(\ref{eksp}) and relevant discussion in the text).}
\label{prof816}
\end{figure}

In order to motivate the results of this subsection, let us first
limit ourselves to what happens in the case of the solutions of
Figures \ref{qtor8} and \ref{qtor16}. Figure \ref{prof816} shows the
average harmonic energy of each mode over a time span $T=10^6$ in
the cases of the $q$-torus of Fig. \ref{qtor8} and Fig.
\ref{qtor16}, shown in Fig. \ref{prof816}a and Fig. \ref{prof816}b
respectively. The numerical result (open circles) compares
excellently with the analytical result (stars) obtained via the
Poincar\'{e} - Lindstedt method. The filled circles in each plot
represent `piecewise' estimates of the localization profile in
groups of modes excited at consecutive orders of the recursive
scheme. The derivation and exact meaning of these theoretical
estimates will be analyzed in detail below. Here we point out their
main feature, showing a clear-cut separation of the modes into {\it
groups} following essentially the sequence of excitations predicted
by the Poincar\'{e}-Lindstedt series construction. Namely, in Fig.
\ref{prof816}a we see clearly that the decrease of the average
energy $\bar{E}_q$ along the profile occurs by abrupt steps, with
three groups formed by nearby energies, namely the group of modes
1,2, then 3 to 6, and then 7,8. The same phenomenon is also seen in
Fig. \ref{prof816}b, where the grouping of the different modes
follows precisely the sequence of excitations (1 to 4), (5 to 12),
(13 to 16) as predicted by the theory.

Let us, therefore, extend our approach to a more general
consideration of the structure of the solutions lying on
low-dimensional tori. Using the Poincar\'{e} - Lindstedt method, a
solution and its frequencies are expanded in series of the form
\begin{eqnarray}\label{P-L}
Q _q(t) = \sum _{k=0}^{\infty } \sigma^kQ_q^{(k)}(t) ,
~~~~\omega _q= \sum _{k=0} ^{\infty} \sigma^{k} \omega _{q} ^{(k)}
\end{eqnarray}
where $\omega_q^{(0)}\equiv\Omega_q$.
Substituting (\ref{P-L}) into the equations of motion (\ref{eqmo})
and separating terms of like orders, the equations at order $k$ read
\begin{eqnarray}\label{mod}
\ddot {Q}_q^{(k)}+\omega _q^2Q_q^{(k)}= \sum _{n_{1}=1}^{k}
(\sum _{n_{2}=0} ^{n_{1}} \omega _{q}^{(n_{2})}
\omega _{q}^{(n_{1} - n_{2})})Q_q^{(k-n_{1})} - \nonumber \\
\Omega _q \sum_{l,m,n=1}^{N}\Omega _l\Omega _m
\Omega _nC_{qlmn}\sum_{\mathop{n_{1,2,3}=0}\limits_
{n_1+n_2+n_3=k-1}}^{k-1}Q_l^{(n_1)}Q_m^{(n_2)}Q_n^{(n_3)}~.
\end{eqnarray}

Note that Eqs.(\ref{mod}) are still quite general. Let us consider,
therefore, the case
where only a subset of modes $1\leq q_1<q_2<...<q_s<N$ are excited
at the zeroth order of the perturbation theory, assuming that
$Q_q^{(0)}\neq 0$ iff $q\in\{q_1,q_2,\ldots,q_s\}$. The modes
$q_1,\ldots,q_s$ need not be consecutive. We wish to see how this
type of zeroth order excitation propagates at subsequent orders.
More specifically, we wish to determine for which $q$ values one has
$Q_q^{(k'<k)}=0$, $Q_q^{(k)}\neq 0$, i.e., the modes $q$ are first
excited at the k-th order. The answer is provided by the
following\\
\\
{\bf Proposition:} {\it Let the starting terms of a
Poincar\'{e}-Lindstedt series solution with $s$ frequencies be set as
\begin{eqnarray}\label{init}
Q^{(0)}_{q_{i}}&=&A_{q_{i}}cos(\omega _{q_{i}} t+\phi_{q_i})~~~
\mbox{for}~i=1,2,...,s,~~1\leq q_1\leq q_2\leq\ldots\leq q_s\leq N\nonumber\\
Q^{(0)}_q&=&0 ~~~ \mbox{for all}~q\neq q_i~~.
\end{eqnarray}
Then, besides the terms $Q_{q_i}^{k}(t)$, the Poincar\'{e}-Lindstedt
series terms $Q_q^{(k)}(t)$ which are permitted to be non-zero at
the $k$th order of the series expansion are given by the values of
$q=q^{(k)}$ satisfying
 \begin{eqnarray}\label{app}
q^{(k)}=\mid 2\lambda(N+1)-m_k \mid,~~~~m_k(mod(N+1))\neq 0
 \end{eqnarray}
where $m_k$ can take any of the values
$\mid {\pm q_{i_1}\pm q_{i_2}\pm ...\pm q_{i_{2k+1}}}\mid$,
with $i_1,i_2,...,i_{2k+1}\in \{1,...,s\}$ for any possible combination
of the $\pm$ signs, and $\lambda\equiv[(m_k+N)/2(N+1)]$.
}\\
\\
The proof of the proposition is given in Appendix A. Some simple
examples clarify the use of the rule (\ref{app}):

{\it $q$--Breathers:} If we excite only one mode $q_1$ at zeroth
order, new modes are excited one by one at subsequent orders and one
has $m_k=(2k+1)q_1$. As long as $m_k \leq N$, the newly excited
modes are $q^{(k)}=m_k=(2k+1)q_1$, i.e., exactly as predicted by
Flach et al. \cite{flaetal2005}. A particular case arises when
$q_1=2(N+1)/3,(N+1)/2$ or $(N+1)/3$. Then, one readily sees that no
new modes are excited at any subsequent order, which is in agreement
with a well known result \cite{bivetal1973} (see also
\cite{antbou2006}).

{\it 2-dimensional $q$--tori:} Assume we excite the modes $q_1=1$,
$q_2=2$ at zeroth order. At first order ($k=1$) we have again
$q=|q_{i_1}\pm q_{i_2}\pm q_{i_3}|$ with $i_1,i_2,i_3\in\{1,2\}$. We
readily find that the newly excited modes are $q=1+1+1=3$,
$q=1+1+2=4$, $q=1+2+2=5$, and $q=2+2+2=6$. At order $k=2$, the first
newly excited mode is $q=1+1+1+2+2=7$, while the last newly excited
mode is $q=2+2+2+2+2=10$. In general, at order $k\geq 1$ the newly
excited modes are $2(2k-1)+1\leq q\leq 2(2k+1)$.

{\it $s$--dimensional $q$--tori}:  Assume we excite the modes
$q_1=1$, $q_2=2$, $\ldots$, $q_s=s$ at the zeroth order. In the same
way as above we find that the newly excited modes at order $k\geq 1$
are $s(2k-1)+1\leq q\leq s(2k+1)$.

More complicated choices of the initially excited modes $q_1,\ldots, q_s$
lead to very interesting localization patterns that will be the subject
of a separate study.

\subsection{Profile of the energy localization}

In order to study now in Fourier space energy localization phenomena
associated with FPU-trajectories, let us construct estimates for all
$Q_q(t)$ terms participating in a particular $q$-torus solution in
which the $s$ first modes $(q_1,q_2,\ldots,q_s)=(1,2,\ldots,s)$ are
excited at zeroth order of the theory, with amplitudes
$A_1,A_2,\ldots,A_s$ respectively.

Denoting by $q^{(k)}$ the indices of all modes which are newly
excited at $k$-th order, according to the proposition of the
previous subsection, one has
\begin{equation}\label{qk}
q^{(k)}\in {\cal M}_{s,k}\equiv \left\{\max(1,(2k-1)s+1),\ldots,
(2k+1)s \right\}~~,
\end{equation}
whence the following useful estimates are obtained
\begin{equation}\label{qomest}
\forall q^{(k)}\in{\cal M}_{s,k},k\geq 1~~~~
q^{(k)}\lesssim (2k+1)s,~~~~
\omega_{q^{(k)}}\lesssim {(2k+1)\pi s\over N+1}~~.
\end{equation}

Let us now define the `majorant' norm in the space of
trigonometric polynomials $f$
\begin{equation}\label{fnorm}
||f|| = \sum_{\mathbf{k}} |A_{\mathbf{k}}|~~
\end{equation}
where $A_{\mathbf{k}}$ are the coefficients of a trigonometric
polynomial with, say (for simplicity), only cosine terms
\begin{equation}\label{ftrig}
f = \sum_{\mathbf{k}} A_{\mathbf{k}}\cos(\mathbf{k}\cdot\mathbf{\phi})~~.
\end{equation}
The quantity (\ref{fnorm}) satisfies all properties of the norm.
Since the series terms of a mode $Q_{q^{(k)}}(t)$ satisfy
$Q_{q^{(k)}}^{(n)}=0$, $\forall n<k$, $Q_{q^{(k)}}^{(k)}\neq 0$,
from Eq.(\ref{mod}) we deduce that the equation determining
$Q_{q^{(k)}}^{(k)}$ reads
\begin{eqnarray}\label{last}
&~&\ddot {Q}_{q^{(k)}}^{(k)}+\omega _{q^{(k)}}^2Q_{q^{(k)}}^{(k)}=\nonumber\\
&=&-\Omega _{q^{(k)}}
\sum_{\mathop{n_{1,2,3}=0}\limits_ {n_1+n_2+n_3=k-1}}^{k-1}
\left(\sum_{(q^{(n_1)},q^{(n_2)},q^{(n_3)})\in{\cal D}_{q^{(k)}}}
\Omega _{q^{(n_1)}}\Omega _{q^{(n_2)}}\Omega _{q^{(n_3)}} Q_{q^{(n_1)}}^{(n_1)}Q_{q^{(n_2)}}^{(n_2)}Q_{q^{(n_3)}}^{(n_3)}\right)
\end{eqnarray}
where
$$
{\cal D}_q=\left\{(q_{j_1},q_{j_2},q_{j_3}):
q\pm q_{j_1}\pm q_{j_2}\pm q_{j_3}=0\right\}~~~.
$$
Assuming that the Lindstedt series are convergent, omission of
explicit reference to higher order terms $Q_{q^{(k)}}^{(k+1)}$,
$Q_{q^{(k)}}^{(k+2)}$, $\ldots$, is justified as long as only an
estimate of the size of the oscillation amplitude of the mode
$q^{(k)}$ is sought.

The average size of the oscillation amplitudes of each mode along an
$s$-dimensional $q$-torus follows now from estimates on the norms of
the various terms appearing in Eq.(\ref{last}). If we denote by
$A^{(k)}$ the mean value of all the norms $||Q_{q^{(k)}}||$, we find
the following estimate:
\begin{equation}\label{akfin}
A^{(k)} = {(Cs)^kA_0^{2k+1}\over 2k+1}
\end{equation}
where $A_0\equiv A^{(0)}$ is the mean amplitude of the oscillations
of all modes excited at the zeroth order of the perturbation theory,
and $C$ is a constant of order O(1). The proof of (\ref{akfin}) is
deferred to Appendix B, in which the analytical estimate $C\simeq
3/2$ is given. In fact, (\ref{akfin}) is a straightforward
generalization of the estimate given in Flach et al.
\cite{flaetal2006} for $q$-breathers, while the two estimates become
identical (except for the precise value of $C$) if one sets $s=1$ in
Eq.(\ref{akfin}), and $q_0=1$ in Flach's $q$-breathers formulae.

Thus, the $q$-tori provide an explanation for the results reported
in \cite{beretal2004} and \cite{deletal1999} and offer a bridge
between the `natural packet' approach and the interpretation of
energy localization for FPU-trajectories based on $q$-breathers. In
particular, the physical picture suggested by the above analysis is
that, starting with initial conditions near $q$-breathers, a
`backbone' is formed in the phase space by a hierarchical set of
solutions which are, precisely, the solutions lying on
low-dimensional $q$-tori (of dimension $s=1,2,..., s<<N$). All
FPU-trajectories with initial conditions within this set exhibit a
profile of the energy localization characterized by a `stepwise'
exponential decay, with step size equal to $2s$, as implied by
Eq.(\ref{qk}). All the modes $q^{(k)}\in {\cal M}_{k,s}$ share a
nearly equal mean amplitude of oscillations, which follows the
estimate of Eq.(\ref{akfin}).

Using (\ref{akfin}), we find it convenient to obtain `piecewise'
estimates of the energy of each group using a formula for the
average harmonic energies $E^{(k)}$ of the modes $q^{(k)}$. To
achieve this, note that the total energy $E$ given to the system
can be estimated as the sum of the energies of the modes
$1,\ldots,s$ (the remaining modes yield only small corrections to
the total energy), i.e.
$$
E\sim s\omega_{q^{(0)}}^2 A_0^2\sim {\pi^2s^3A_0^2\over (N+1)^2}~~.
$$
On the other hand, the energy of each mode $q^{(k)}$ can be estimated from
$$
E^{(k)}\sim {1\over 2}\Omega_{q^{(k)}}^2
\left({\beta\over 2(N+1)}\right)^{2k}(A^{(k)})^2\sim
{\pi^2s^2(Cs\beta)^{2k}A_0^{4k+2}\over 2^{2k+1}(N+1)^{2k+2}}
$$
which, in terms of the total energy $E$, yields
\begin{equation}\label{ek}
E^{(k)}\sim {E\over s}\left({C^2\beta^2(N+1)^2E^2\over \pi^4s^4}\right)^k~~.
\end{equation}
Once again, the similarity of Eq.(\ref{ek}) with the corresponding
equation for $q$-breathers is obvious. The latter equation reads
\cite{flaetal2006}
\begin{equation}\label{ekqb}
E_{(2k+1)q_0}\sim E_{q_0}\left({9\beta^2(N+1)^2E^2\over 64 \pi^4q_0^4}\right)^k
\end{equation}
where $q_0$ is the unique mode excited at zeroth order of the
perturbation theory. Note, in particular, that the integer $s$ plays
in Eq.(\ref{ek}) a role similar to that of $q_0$ in Eq.(\ref{ekqb}).
This means that the energy profile of a $q$-breather with $q_0=s$
presents the same exponential law as the energy profile of the
$s$-dimensional $q$-torus. But the most important feature of the
latter type of solutions is that the profile remains {\it unaltered}
as $N$ increases, provided that: i) {\it a constant fraction}
$M=s/N$ of the spectrum is initially excited, (i.e. that $s$
increases proportionally to $N$), and ii) the {\it specific energy
$\varepsilon=E/N$ remains constant}. Indeed, in terms of the
specific energy $\varepsilon$, (\ref{ek}) takes the form
\begin{equation}\label{eksp}
E^{(k)}\sim {\varepsilon\over M}
\left({C^2\beta^2\varepsilon^2\over \pi^4M^4}\right)^k~~
\end{equation}
i.e. the profile becomes {\it independent} of $N$. A similar
behavior is recovered in the $q$-breather solutions provided that
the `seed' mode $q_0$ varies linearly with $N$, as was shown in
detail in refs.\cite{flaetal2007,kanetal2007}.

\subsection{Numerical results for FPU-trajectories}

Examples of the `stepwise' profiles predicted by Eq.(\ref{ek}) in
the case of exact $q$-tori solutions are shown by filled circles in
Fig. \ref{prof816}, concerning the solutions depicted in Figs.
\ref{qtor8} and \ref{qtor16} (in all fittings we set $C=1$ for
simplicity). From these one can see that the theoretical `piecewise'
profiles yield nearly the same average exponential slope as the
profiles obtained either numerically, or analytically by the
construction of the solutions via the Poincar\'{e} - Lindstedt
method. Thus, the estimates (\ref{ek}), or (\ref{eksp}), appear
quite satisfactory for characterizing the localization profiles of
exact $q$-tori solutions.

The key question now, regarding the relevance of the $q$-tori
solutions for the interpretation of the FPU paradox, is whether
Eqs.(\ref{ek}) or (\ref{eksp}) retain their predictive power in the
case of generic FPU trajectories which, by definition, are
trajectories started close to, but not exactly on a $q$-torus.

\begin{figure}
\centering
\includegraphics[scale=0.6]{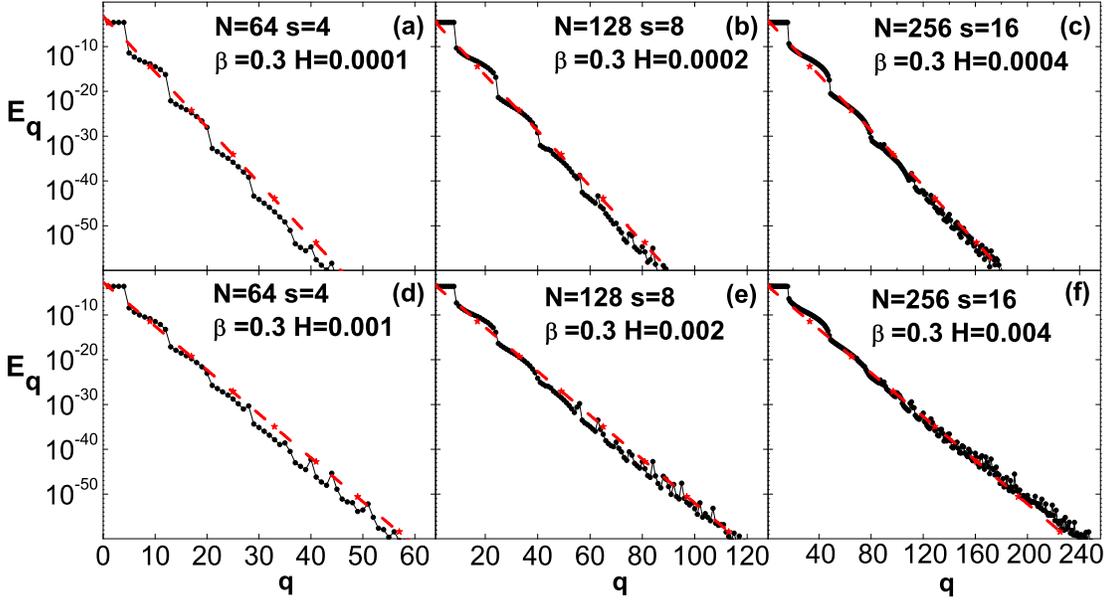}
\caption{(Color
 online) The average harmonic energy $E_q$ of the $q$-th mode over a
time span $T=10^6$ as a function of $q$ in various examples of
FPU-trajectories, for $\beta=0.3$, in which the $s(=N/16)$ first
modes are only excited initially via $Q_q(0)=A_q$, $\dot{Q}_q(0)=0$,
$q=1,\ldots,s$, with the $A_q$ selected so that the total energy is
equal to the value $E=H$ indicated in each panel. We thus have (a)
$N=64$, $E=10^{-4}$, (b) $N=128$, $E=2\times 10^{-4}$, (c) $N=256$,
$E=4\times 10^{-4}$, (d) $N=64$, $E=10^{-3}$, (e) $N=128$,
$E=2\times 10^{-3}$, (f) $N=256$, $E=4\times 10^{-3}$. The specific
energy is constant in each of the two rows, i.e.
$\varepsilon=1.5625\times 10^{-6}$ in the top row and
$\varepsilon=1.5625\times 10^{-5}$ in the bottom row. The dashed
lines represent the average exponential profile $E_q$ obtained
theoretically by the hypothesis that the depicted FPU trajectories
lie close to $q$-tori governed by the profile (\ref{eksp}).}
\label{profpu1}
\end{figure}
\begin{figure}
\centering
\includegraphics[scale=0.6]{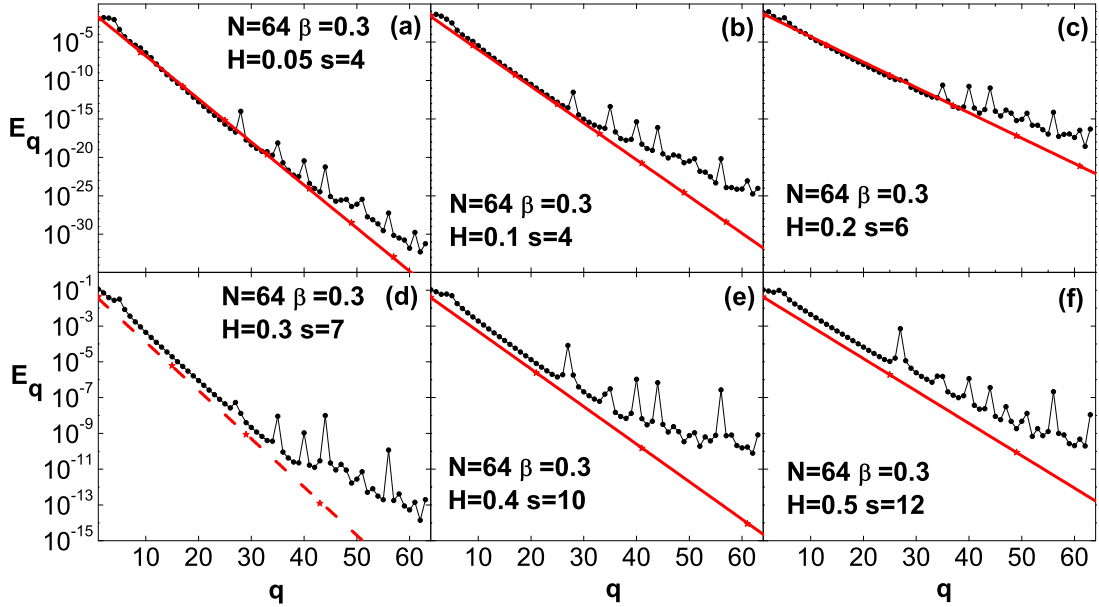}
\caption{(Color
 online) Same as in Fig.\ref{profpu1}a but for larger energies, namely
(a) $E=0.05$, (b) $E=0.1$, (c) $E=0.2$, (d) $E=0.3$, (e) $E=0.4$,
(f) $E=0.5$. Beyond the threshold $E\simeq 0.05$, theoretical
profiles of the form (\ref{ek}) yield the correct exponential slope
if $s$ is gradually increased from $s=4$  in (a) and (b) to $s=6$ in
(c), $s=7$ in (d), $s=10$ in (e), and $s=12$ in (f). }
\label{profpu2}
\end{figure}

An answer to this question is partly contained in the results
presented in Figures \ref{profpu1} and \ref{profpu2}. Figure
\ref{profpu1} shows the energy localization profile in numerical
experiments in which $\beta$ is kept fixed ($\beta=0.3$), while $N$
takes the values $N=64$, $N=128$ and $N=256$ (although in subsection
IIB the derivation of explicit $q$-tori solutions was practically
feasible by computer algebra only up to a rather small value of $N$
($N=16$), in the present subsection the results with numerical
trajectories are extended to much higher values of $N$). In all six
panels of Fig. \ref{profpu1} the FPU-trajectories are computed
starting with initial conditions in which {\it only} the $s=4$ (for
$N=64$), $s=8$ (for $N=128$) and $s=16$ (for $N=256$) first modes
are excited at $t=0$, with the excitation amplitudes being
compatible with the values of the total energy $E$ indicated in each
panel, and constant specific energy $\varepsilon=1.5625\times
10^{-6}$ in the top row and $\varepsilon=1.5625\times 10^{-5}$ in
the bottom row of Fig. \ref{profpu1}.

The resulting trajectories differ from $q$-tori solutions as
follows: In the $q$-tori all modes have an initial excitation, whose
size was estimated in Eqs.(\ref{akfin}) or (\ref{ek}), while in the
case of the FPU trajectories only the $s$ first modes are excited
initially, and one has $Q_q(0)=0$ for all modes $q>s$. Furthermore,
since in the $q$-tori solutions one also has
$||Q_{q>s}(t)||<<||Q_{q\leq s}(t)||$ for all $t$, the FPU
trajectories can be considered as lying in the neighborhood of the
$q$-torus solutions, at least initially. The numerical evidence is
that if $E$ is small, they remain close to the $q$-tori even after
relatively long times, e.g. $t=10^6$.

This is exemplified in Fig. \ref{profpu1}, in which one sees that
the average energy profiles of the FPU-trajectories (at $t=10^6$)
exhibit the same behavior as predicted by Eq.(\ref{ek}), for an
exact $q$-torus solution with the same total energy as the FPU
trajectory in each panel. For example, based on the values of their
average harmonic energy, the modes in Fig. \ref{profpu1}a (in which
$s=4$) are clearly separated in groups, (1 to 4), (5 to 12) and (13
to 20), etc., as foreseen by Eq.(\ref{qk}) for an exact 4-torus
solution. The energies of the modes in each group have a sigmoid
variation around a level value characteristic of the group, which is
nearly the value predicted by Eq.(\ref{ek}). The grouping of the
modes is distinguishable in all the panels of Fig. \ref{profpu1}, a
careful inspection of which verifies that the grouping follows the
laws found above for $q$-tori. Also, if we superpose the numerical
data of the three top (or bottom) panels we find that the average
exponential slope is nearly identical in all the panels of each row,
a fact consistent with Eq.(\ref{eksp}), according to which, for a
given fraction $M$ of initially excited modes, this slope depends on
the specific energy only, i.e. is independent of $N$ for constant
$\varepsilon$.

When we increase the energy, the FPU-trajectories resulting from $s$
initially excited modes start deviating from their associated exact
$q$-tori solutions. As a consequence, the energy profiles of the
FPU-trajectories start also deviating from the energy profiles of
the exact s-tori. This is evidenced by the fact that the profiles of
the FPU-trajectories become smoother, and the groups of modes less
distinct, while retaining the average exponential slope predicted by
Eq.(\ref{eksp}). This `smoothing' of the profiles is discernible in
Fig. \ref{profpu2}a, in which the energy is increased by a factor 50
with respect to Fig. \ref{profpu1}d, for the same values of $N$ and
$\beta$. Also, in Fig. \ref{profpu2}a we observe the formation of
the so-called `tail', i.e. an overall rise of the localization
profile at the high-frequency part of the spectrum, accompanied by
spikes at particular modes. This is a precursor of the evolution of
the system towards equipartition, which manifests itself earlier in
time for larger energies.

Nevertheless, the important remark is that the phenomenon of
exponential localization of the FPU trajectories persists, and is
still characterized by laws like Eq.(\ref{ek}), even when the
energy is substantially increased. Furthermore, at energies beyond
a threshold value, an interesting phenomenon occurs which is worth
mentioning: For fixed $N$ (see e.g. Figure \ref{profpu2}, where
$N=64$), as the energy increases, a progressively higher value of
$s$ needs to be used in Eq.(\ref{ek}), so that the theoretical
profile yields an exponential slope that agrees with the numerical
data.

As evidenced in Fig. \ref{profpu2} for $\beta=0.3$, $N=64$, the
threshold is $E\approx 0.1$. This value splits the system in two
distinct regimes: One for $E<0.1$, where the numerical data are well
fitted by a constant choice of $s=4$ in Eq.(\ref{ek}) (indicating
that the FPU-trajectories are indeed close to 4-tori), and another
for $E>0.1$, where `best-fit' models of Eq.(\ref{ek}) occur for
values of $s$ increasing with the energy, e.g. $s=6$ for $E=0.2$,
rising to $s=12$ for $E=0.5$. This indicates that the respective
FPU-trajectories are close to $q$-tori with a progressively higher
value of $s$ (with $s>4$), despite the fact that only the four first
modes are excited by the initial conditions of these trajectories.

This behavior is analogous to the `natural packet' scenario
described by Berchialla et al. \cite{beretal2005}, in which a set of
modes is seen to share the energy after some time even if this
energy is  initially given to only the first mode. These authors
also observed that the law giving the localization profile of their
metastable states stabilizes as the energy increases. Indeed,
according to Eq.(\ref{ek}), such a stabilization implies that in the
second regime the width $s$ depends asymptotically on $E$ as
$s\propto E^{1/2}$, or, from Eq.(\ref{eksp}) that $M\propto
\varepsilon^{1/2}$. This agrees well with estimates on the width of
natural packets formed by the $\beta-$FPU model described in
\cite{lichetal2008}.

\section{Stability of the motion near $q$-tori}

\subsection{Linear stability}
The linear stability of $q$-breathers can be studied by the
implementation of Floquet theory (see \cite{flaetal2006}), which
demonstrates that a $q$-breather is linearly stable as long as
\begin{equation}\label{qbst}
R={6\beta E_{q_0}(N+1)\over \pi^2}<1+O(1/N^2)
\end{equation}
This result is obtained by analyzing the eigenvalues of the
monodromy matrix of the linearized equations about a $q$-breather
solution constructed by the Poincar\'{e} - Lindstedt series.
Numerical verification can also be used to analyze the dynamics
about the fixed point corresponding to a $q$-breather under the
Poincar\'{e} map of the flow of Eqs.(\ref{eqmo}).

In the case of $q$-tori the above techniques are no longer
available. Nevertheless, a reliable numerical criterion for the
stability of $q$-tori is provided by the GALI indices
\cite{skoetal2007}. According to this method, if a $q$-torus becomes
unstable beyond a critical energy threshold, the deviation vectors
of trajectories started exactly on the $q$-torus are attracted by
its unstable manifold, whence all the GALI indices beyond and
including $GALI_{k+1}$ ($k$ being the dimension of the unstable
manifold) fall exponentially fast. Naturally, if one starts with
trajectories in the vicinity of an unstable $q$-torus, these
trajectories are weakly chaotic. Thus, it turns out that all GALI
indices start falling exponentially after a transient time, and this
can be checked by calculating the time evolution of the lowest
index, i.e. $GALI_2$.

\begin{figure}
\includegraphics[scale=0.3]{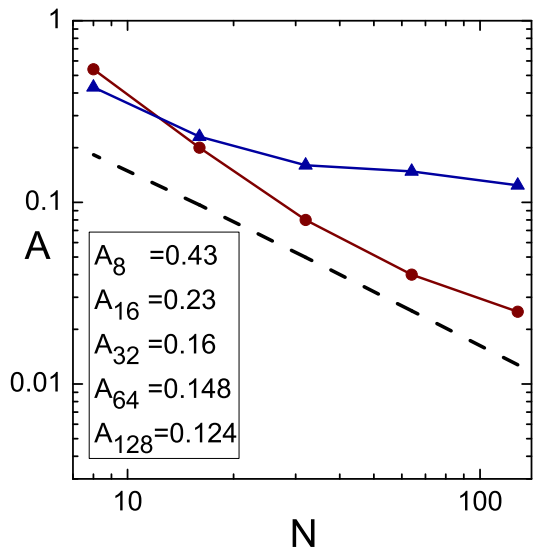}
\caption{(Color
 online) Assuming that the critical energy $E_c$, at which the
$GALI_2$ index shows that a $q$-torus destabilizes is fitted by the
law $E_c= A \beta^{-1}$, we show in the upper curve (triangles) the
dependence of $A$ on $N$ for an FPU trajectory started by exciting
initially only the $q=1,2$ modes. The middle curve (filled circles)
corresponds to a similar calculation for FPU trajectories started
near a $q$-breather solution, where only the $q=1$ mode is excited.
The dashed line corresponds to $A\sim N^{-1}$, according to the law
of Eq.(\ref{qbst}).} \label{linsta}
\end{figure}
Using this criterion, we examined the stability of $q$-tori and
determined approximately the value of the critical energy $E_c$ at
which a $q$-torus turns from stable to unstable. Figure \ref{linsta}
shows an example of this calculation using FPU trajectories started
close to a 2-torus. The quantity $A$ shown in the ordinate
corresponds to a calculation keeping $N$ fixed and varying $\beta$,
until a critical energy $E_c$ is determined, beyond which the
$GALI_2$ index loses its asymptotically constant behavior. All
calculations refer to a maximum time $t_{max}=10^7$ up to which we
require that the exponential fall-of of $GALI_2$ must have been
observed (in general $E_c$ decreases as $t_{max}$ increases, tending
to an asymptotic limit as $t_{max}\rightarrow\infty$). The values of
$E_c$ found this way provide {\it upper estimates} for the
transition energy at which the exact $q$-torus turns from stable to
unstable, i.e., for any choice of $t_{max}$ the transition energy is
{\it lower} than the value of $E_c$ found by the $GALI_2$ method.

As expected, due to the obvious scaling of the FPU Hamiltonian by
$\beta$, the critical energy for all considered values of $N$ turns
out to be well fitted by a power law $E_c= A \beta^{-1}$. However,
the fitting constant $A$ depends also on $N$, as seen in Fig.
\ref{linsta}. From the triangle data in the figure it is clear that
the dependence of $A$ on $N$ is weaker than $N^{-1}$, i.e. the law
predicted by (\ref{qbst}) for the $q$-breathers.

On the other hand, the dependence $A\propto N^{-1}$ was
approximately found for FPU-trajectories started close to a
$q$-breather, where only the mode $q=1$ is initially excited. In
that case the numerical data (filled circles) have a slope close to
that predicted by Eq.(\ref{qbst}), although the whole numerical
curve is shifted upwards with respect to the dashed line, a fact
verifying that the $GALI_2$ method yields critical energies $E_c$
which are \textit{higher} than the value at which the $q$-breather
becomes unstable.

\begin{figure}
\includegraphics[scale=0.45]{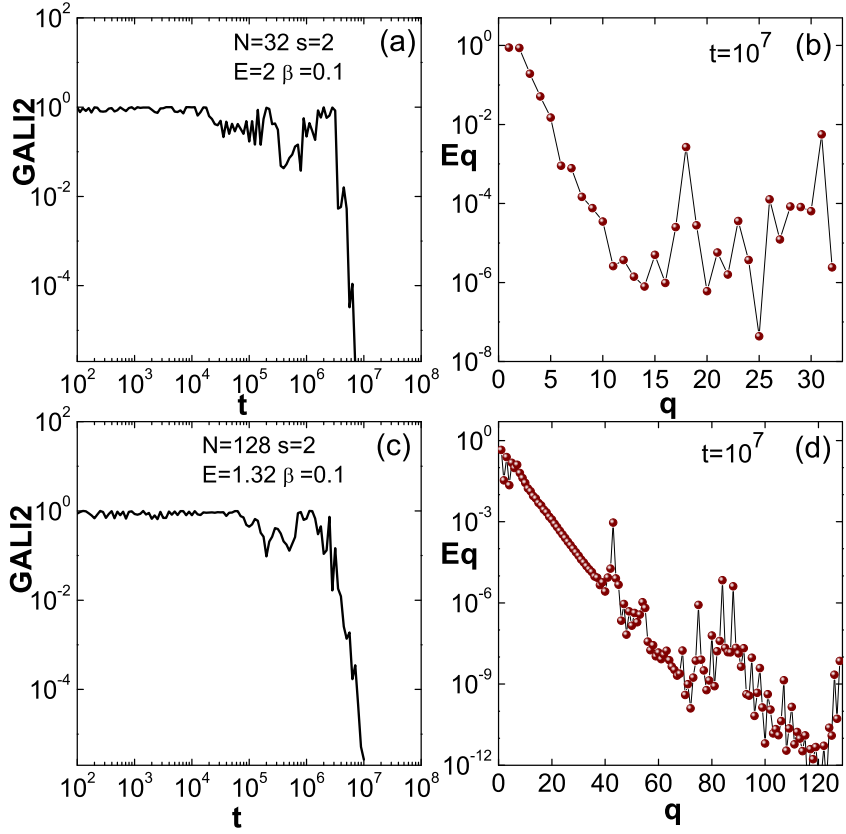}
\caption{(Color
 online) (a) Time evolution of the $GALI_2$ index up to $t=10^7$ for an
FPU trajectory started by exciting the $q=1$ and $q=2$ modes in the
system $N=32$, $\beta=0.1$, with total energy $E=2$, i.e. higher
than $E_c=1.6$. (b) {\it Instantaneous} localization profile of the
FPU trajectory of (a) at $t=10^7$. (c) Same as in (a) but for
$N=64$, $E=1.323$ (in this case the critical energy is $E_c=1.24$).
(d) Same as in (b) but for the trajectory of (c).} \label{linsta2}
\end{figure}
The weak dependence of $A$ on $N$ on the upper curve of Fig.
\ref{linsta} is a numerical indication that the $q$-tori solutions
are \textit{more robust} than the $q$-breathers regarding their
linear stability. This numerical behavior is related to the results
of the previous subsection,  which indicate that the number of modes
excited initially does not always coincide with the dimension of the
$q$-torus which the FPU-trajectory eventually approaches. Also, the
destabilization of simple periodic orbit represented by a
$q$-breather does not imply that the tori surrounding the breather
are also be unstable.

At any rate, the most important remark concerning the linear
stability of $q$-breathers or $q$-tori is that the exponential
localization of FPU trajectories persists even after the associated
$q$-breathers or $q$-tori have been identified as linearly unstable
by the GALI criterion. This behavior is exemplified in Figure
\ref{linsta2}, where panels (a) and (c) show the time evolution of
the $GALI_2$ index for two FPU trajectories started in the vicinity
of 2-tori of the $N=32$ and $N=128$ systems, when $\beta=0.1$ and
$t_{max}=10^7$. In both cases, the energy satisfies $E>E_c$, as the
exponential fall-off of the $GALI_2$ index is already observed at
$t=10^7$. However, a simple inspection of Figs. \ref{linsta2}b,d
clearly reveals that the exponential localization of the energy
persists in the Fourier space of both systems. In fact we have found
that the exponential localization persists for energies much larger
than $E_c$, but for timescales which become smaller as $E$
increases.

\subsection{A heuristic argument on exponential stability}

The results of the previous subsection confirm the observation made
on the basis of numerical experiments in \cite{beretal2004},
that a result establishing \textit{exponential stability} for FPU
trajectories close to $q$-tori should be possible. This has already
been discussed in the introduction, where we mentioned a partial
result in this direction obtained recently in \cite{giomur2006}.
It was pointed out, however, that so far the main obstruction to
precise analytical statements lies in the bad dependence of all
existing estimates on $N$.

Thus, in closing our paper, we would like to offer a heuristic
argument on how the property of {\it exponential localization of
the energy} for particular FPU trajectories could
serve as a basis for further improvement of rigorous results. For
definiteness, we refer below to the case of FPU-trajectories for
which the question of exponential stability can be examined in the
framework of a variant of Birkhoff's method due to Giorgilli
\cite{gio1988}, based on the direct calculation of approximate
integrals of motion without the use of normal forms \cite{whi1916,
che1924, con1960}.

The trajectories we are referring to have initial conditions close
to a particular family of $q$-breather solutions considered in
\cite{flaetal2007, kanetal2007}, whose `seed mode' $q_0$ varies
proportionally to $N$. We thus take the seed mode to vary as
$q_0=(N+1)/2\mu$, where $\mu$ can be any large number that is a
power of two, while $N$ is a power of two minus 1, so that no
commensurabilities exist in the set of $N$ unperturbed FPU
frequencies $\Omega_q$, $q=1,...N$. The exact $q$-breather solution
oscillates with just one frequency and thus lies on a
one-dimensional torus. However, perturbing this trajectory, we
obtain solutions lying on $\mu-$dimensional tori involving the modes
$q=(2j-1)q_0$, $j=1,2,...,\mu$. Furthermore, the energy localization
profile for the $q$-breather solution is the same as that of a
$s$-dimensional $q$-torus solution with $s=q_0$ given by
Eq.(\ref{eksp}), where $M=q_0/N\simeq 1/2\mu$.

Let us assume that this localization profile holds also initially
for the FPU-trajectory that is a perturbation of the exact
$q$-breather solution. Our aim then is to demonstrate that when one
establishes estimates of exponential time of stability
$T\sim\exp(1/\varepsilon^c)$ for the energy localization profile of
this type of FPU-trajectories, the exponent $c$ does not depend on
the number of degrees of freedom (i.e. is independent of $\mu$),
despite the diophantine character of the unperturbed frequencies
$\Omega_q$ of the $\mu$ modes $q=(2j-1)q_0$, $j=1,2,...,\mu$.

To this end, we recall that, according to \cite{gio1988}, the
question of the long-term stability of particular trajectories can
be examined by constructing approximate integrals of motion in the
form of truncated series starting with harmonic energies
\begin{equation}\label{phij}
\Phi_j=\Phi_j^{(2)}+\Phi_j^{(4)}+...
\end{equation}
where $\Phi_j^{(r)}$ is an $r$th degree polynomial in the canonical
variables $(q_1,\ldots,q_N,p_1,\ldots,p_N)$ defined by
\begin{equation}\label{cqp}
q_j={\sqrt{\Omega_j}Q_j-iP_j/\sqrt{\Omega_j}\over \sqrt{2}},~~~
p_j={-i\sqrt{\Omega_j}Q_j+P_j/\sqrt{\Omega_j}\over \sqrt{2}}
\end{equation}
and $\Phi_j^{(2)}\equiv E_j=i\Omega_jp_jq_j$. The functions
$\Phi_j^{(r)}$ are determined recursively, by solving formally the
Poisson bracket condition
$\{\Phi_j^{(2)}+\Phi_j^{(4)}+...,H_2+H_4\}=0$, demanding that the
series $\Phi_j$ be an integral. Here $H_2$ and $H_4$ are the
quadratic and quartic terms of the $\beta$-FPU Hamiltonian written
in the variables $(q_j,p_j)$. The order by order solution is then
found via the so-called homological equation
\begin{equation}\label{homo}
\{\Phi_j^{(r-2)},H_4\}+\{\Phi_j^{(r)},H_2\}=0,~~~r=4,6,...
\end{equation}
which has the same algebraic structure as the homological equation of
Birkhoff's normalization scheme. The solution reads
\begin{equation}\label{homosol}
\Phi_j^{(r)} = -\sum_{\mathop{m,n=1}\limits_{m+n=r}}^r
\frac{h_{m,n}^{(r)}\mathbf{q}^\mathbf{m}\mathbf{p}^\mathbf{n}}
{(\mathbf{m}-\mathbf{n})\cdot\mbox{\boldmath$\Omega$}}
\end{equation}
where use is made of the compact notation
$\mathbf{q}^\mathbf{m}\mathbf{p}^\mathbf{n}\equiv
q_1^{m_1}q_2^{m_2}\ldots q_N^{m_N}p_1^{n_1}p_2^{n_2}\ldots
p_N^{m_N}$, $\mathbf{m}\equiv(m_1,m_2,\ldots,m_N),~~~m\equiv
m_1+m_2+\ldots+m_N$, (similarly for $\mathbf{n}$), and
$h_{m,n}^{(r)}$ are the polynomial coefficients of the Poisson
bracket $\{\Phi_j^{(r-2)},H_4\}$.

It is well known that the series (\ref{phij}) is asymptotic. We thus
consider the $r$th order finite truncation
$\Phi_{j,r}=\Phi_j^{(2)}+\Phi_j^{(4)}+\ldots+\Phi_j^{(r)}$
representing an {\it approximate} integral of motion whose time
variation is given by
\begin{equation}\label{remj}
{d\Phi_{j,r}\over dt}=R_{j,r}\equiv\left\{\Phi_j^{(r)},H_4\right\}~~.
\end{equation}
The quantity $R_{j,r}$ is called the remainder function. The
asymptotic character of the series implies that the size of
$R_{j,r}$ decreases initially as $r$ increases, up to an optimal
order $r=r_{opt}$ beyond which the size of the remainder increases
with $r$, becoming ultimately divergent as $r\rightarrow\infty$.
Recursive application of the homological equation (\ref{homo})
(see \cite{gio1988}, \cite{eftetal2004} for details) yields that
the size of the remainder at order $r$ can be estimated by
\begin{equation}\label{remrec}
||R_{j,r}||\sim{(r-2)||R_{j,r-2}||\over a_{r-2}}
\sim{(r-2)(r-4)||R_{j,r-4}||\over a_{r-2}a_{r-4}}\sim\ldots
\sim{(r-2)!!||R_{j,4}||\over a_{r-2}a_{r-4}...a_{4}}
\end{equation}
where $||R_{j,r}||$ denotes the absolute sum of the polynomial
coefficients $h_{m,n}$ for all $(\mathbf{m},\mathbf{n})$ with
$|\mathbf{m}|+|\mathbf{n}|=r$ and $a_k$, $k=4,6,...,r-2$, denotes
the minimum of all divisors
$(\mathbf{m}-\mathbf{n})\cdot\mbox{\boldmath$\Omega$}$,
$|\mathbf{m}|+|\mathbf{n}|=k$ at order $k$. The pre-factors
$(r-2),(r-4)...$ come from the derivatives of the functions
$\Phi_j^{(r-2)}$, $\Phi_j^{(r-4)}$,..., with respect to the
canonical variables $(\mathbf{q},\mathbf{p})$, appearing in Poisson
brackets due to recursive application of the homological equation,
which yield factors $|\mathbf{m}|,|\mathbf{n}|$, both of order
$O(r-2)$, $O(r-4)$,... within the functions $\Phi_j^{(r-2)}$,
$\Phi_j^{(r-4)}$,..., $\Phi_j^{(4)}$.

Our heuristic argument now goes as follows: Since in the
considered trajectories the total number of participating modes
is equal to $\mu$, and $N$ is a power of two minus one,
the associated divisors satisfy a {\it diophantine condition}
of the form
\begin{equation}\label{dioph}
|\mathbf{l}\cdot\mbox{\boldmath$\Omega$}|\geq{\gamma\over |\mathbf{l}|^\tau}~~~
\mbox{for all}~\mathbf{l}\in\mathbb{Z}^N, |\mathbf{l}|\neq 0
\end{equation}
with $\tau>\mu-1$. The estimate $\tau\sim\mu$ holds for $\mu$ large.
One then has $a_{r-2}a_{r-4}...a_{4}$ $\sim \gamma^{r-4\over
2}(r-2)!!^{-\mu}$, which, upon substitution in (\ref{remrec}) leads
to:
\begin{equation}\label{remrec3}
||R_{j,r}||\sim(r-2)!!^{\mu+1}{\cal C_*}^r\left(\beta\over N\right)^{r/2}~~
\end{equation}
where the estimate $||H_4||=O(\beta/N)$ is taken into account and
${\cal C}_*$ is a $O(1)$ positive constant. These estimates, based
on small divisors, are standard and lead by themselves to no
improvement as far as the dependence of the asymptotic character on
$\mu$ is concerned. However, an improvement can be achieved if we
also take into account the {\it numerators} of the remainder series,
which for FPU-trajectories possess the exponential localization
profile shown in Eq.(\ref{ekqb}). In other words, the size of the
remainder depends also on the size of the monomials
$$
\Delta_{\mu}=(\Omega_1^{1/2}\xi_1)^{s_1} (\Omega_2^{1/2}\xi_2)^{s_2}
\ldots (\Omega_\mu^{1/2}\xi_\mu)^{s_\mu}
$$
where $\xi_k$ stands for either $q_k$ or $p_k$, since each term of
the remainder consists of one such monomial multiplied by a
coefficient bounded by an estimate of the form (\ref{remrec3}). In
view of Eq.(\ref{cqp}), one has $|\Omega_k^{1/2}\xi_k|\sim
E_k^{1/2}$. Thus, taking into account the form of the profile
(\ref{ekqb}), the size of the above monomial at the $r$th order of
normalization can be estimated as:
$$
|\Delta_{\mu}|\sim (E_1)^{s_1\over 2}(E_2)^{s_2\over
2}...(E_\mu)^{s_\mu\over 2} \sim E^{s_1+s_2+...+s_\mu\over 2}
\left(\beta\mu^2\varepsilon\right)^{s_2+2s_3+\ldots+(\mu-1)s_\mu}
$$
where $s_1+s_2+\ldots+s_\mu=r$. For the leading terms (with
smallest divisors) in the series the exponents $s_1,s_2,...,s_\mu$
typically take values such that the estimate
$s_2+2s_3+\ldots+(\mu-1)s_\mu\sim \mu r/2$ holds. Thus
\begin{equation}\label{monsize}
|\Delta_{\mu}| \sim {\cal D_*}^rE^{r/2}\varepsilon^{\mu r/2}~~
\end{equation}
where ${\cal D_*}$ is a new constant. Combining now the estimates
(\ref{monsize}) and (\ref{remrec3}) yields a final estimate for
the size of the remainder of the form
\begin{equation}
||R_r||\sim (r-2)!!^{\mu+1}{\cal C_*}^r{\cal D_*}^r
\left(\beta\over N\right)^{r/2}E^{r/2}\varepsilon^{\mu r/2}\sim
|R_r||\sim {\cal B_*}((r-2)!)^{\mu+1\over
2}\varepsilon^{(\mu+1)r/2}~~,\label{remrecfin}
\end{equation}
The optimal order of truncation is found by taking the logarithm
of (\ref{remrecfin}), using Stirling's formula $\log n!\approx
n\log n - n$, as well as $r-2\approx r$ (for $r$ large), using
$d\log ||R_r||/dr \sim 0$. We thus find $r_{opt}\sim
1/\varepsilon$, whence the optimal value of the remainder is
\begin{equation}\label{remopt}
||R_{opt}||\sim
\exp\left(-{\mu+1\over 2\varepsilon}\right)
\end{equation}
i.e. the time variations of the integrals $\Phi_{j,r_{opt}}$ are
exponentially small in $1/\varepsilon$. This yields an estimate of
exponential stability of the form $T\sim\exp(1/\varepsilon^c)$,
where $c=1$, i.e. the number of degrees of freedom no longer appears
in the exponent $c$. Compared to general estimates yielding $c\sim
1/\mu$, it is seen that the removal of the dependence of $c$ on
$\mu$ was possible thanks to the assumed \textit{exponential
scaling} of the energy localization profile, that allows for the
estimate (\ref{monsize}). In turn, since this localization profile
is the same as for $q$-tori, the above analysis is suggestive of the
usefulness of exponential localization in $q$-space in order to
obtain improved estimates in the case of $q$-tori solutions as well.
However, in the lack of a rigorous demonstration, it remains an open
question whether such strategy can lead to estimates of a stretched
exponential law for the dependence of $||R_{opt}||$ on
$\varepsilon$, as evidenced also to a limited extent by the
numerical experiments of \cite{beretal2004}.

\section{Conclusions}
The main conclusions of the present study can be summarized as
follows:

1) We introduced the concept of $q$-tori in FPU lattices, which
represent a generalization of the concept of $q$-breathers. The
$q$-tori have low dimensionality $s<<N$, and arise from the
continuation of motions with $s$ independent frequencies of the
unperturbed problem.

2) We explicitly calculated FPU solutions lying on $q$-tori by
employing the method of Poincar\'{e}-Lindstedt series. Based on
estimates of the leading terms of these series, we provided a
theoretical law yielding the average exponential localization of the
energy in Fourier space for solutions on $q$-tori. Furthermore, a
proposition was proved which explains how different groups of modes
are excited at consecutive orders of the perturbation theory. The
most important conclusion from this analysis is that, if the
fraction $s/N$ is kept constant, it appears that the localization
profile depends on the {\it specific energy} $\varepsilon=E/N$ of
the system, i.e., is independent of $N$.  Some numerical evidence is
provided in support of the above theoretical analysis, but further
numerical work is necessary, in order to clarify the extent of its
validity.

3) We explored numerically the relevance of $q$-tori of dimension
$s$ to the dynamics of FPU-trajectories started nearby, by exciting
$s$ modes only. The localization laws found analytically for
$q$-tori accurately describe the localization of energy in Fourier
space for the FPU-trajectories as well. We also gave numerical
evidence of the existence of two regimes separated by a critical
energy value. Below this energy, the localization profiles
$E(q)\propto\exp(-bq)$ have a slope $b$ depending logarithmically on
$\varepsilon$, while the fraction $M=s/N$ of modes sharing the
energy is constant. Furthermore, beyond this critical energy the
slope of the localization profile tends to stabilize, and $M$ tends
asymptotically to the law $M\propto \varepsilon^{1/2}$, as $q$-tori
of progressively higher dimension begin to describe the dynamics of
the numerical FPU-trajectories.

4) We examined the stability of $q$-tori using a numerical criterion
provided by the GALI indices \cite{skoetal2007}, and provided
numerical evidence demonstrating that the localization in Fourier
space persists for energies well above the threshold value at which
the underlying $q$-tori turn from stable to unstable.

5) Finally, we provided a heuristic argument suggesting that the
exponential energy localization in $q$-mode space implies a
particular analytical structure of the Birkhoff series obtained for
associated FPU-trajectories, that could lead to improved estimates
of the long term stability in the spirit of Nekhoroshev theory. In
the example of FPU trajectories started close to $q$-breather
solutions exhibiting the same energy localisation as for $q$-tori,
our arguments suggest that in stability estimates of the form
$T\sim\exp(1/\varepsilon^c)$ it is possible to remove the bad
dependence of $c$ on the number of degrees of freedom of the
problem. Of course, further study, substantiated by numerical
experiments, is needed before rigorous statements are available on
this issue.

\begin{acknowledgments}

We wish to thank Drs. S. Flach, A. Ponno and A. Giorgilli for
useful discussions clarifying particular points of the paper, as
well as one referee who pointed out some inconsistencies in our
first version. H. Christodoulidi was supported in part by a grant
from I.K.Y., the Foundation of State Scholarships of Greece, and
gratefully acknowledges the hospitality of the Max Planck
Institute for the Physics of Complex Systems during the period of
November 2008 to March 2009.

\end{acknowledgments}


\begin{thebibliography}{42}
\expandafter\ifx\csname natexlab\endcsname\relax\def\natexlab#1{#1}\fi
\expandafter\ifx\csname bibnamefont\endcsname\relax
  \def\bibnamefont#1{#1}\fi
\expandafter\ifx\csname bibfnamefont\endcsname\relax
  \def\bibfnamefont#1{#1}\fi
\expandafter\ifx\csname citenamefont\endcsname\relax
  \def\citenamefont#1{#1}\fi
\expandafter\ifx\csname url\endcsname\relax
  \def\url#1{\texttt{#1}}\fi
\expandafter\ifx\csname urlprefix\endcsname\relax\def\urlprefix{URL }\fi
\providecommand{\bibinfo}[2]{#2}
\providecommand{\eprint}[2][]{\url{#2}}

\bibitem[{\citenamefont{Flach et~al.}(2005)\citenamefont{Flach, Ivanchenko, and
  Kanakov}}]{flaetal2005}
\bibinfo{author}{\bibfnamefont{S.}~\bibnamefont{Flach}},
  \bibinfo{author}{\bibfnamefont{M.~V.} \bibnamefont{Ivanchenko}},
  \bibnamefont{and} \bibinfo{author}{\bibfnamefont{O.~I.}
  \bibnamefont{Kanakov}}, \bibinfo{journal}{Phys. Rev. Lett.}
  \textbf{\bibinfo{volume}{95}}, \bibinfo{pages}{064102}
  (\bibinfo{year}{2005}).

\bibitem[{\citenamefont{Flach et~al.}(2006)\citenamefont{Flach, Ivanchenko, and
  Kanakov}}]{flaetal2006}
\bibinfo{author}{\bibfnamefont{S.}~\bibnamefont{Flach}},
  \bibinfo{author}{\bibfnamefont{M.~V.} \bibnamefont{Ivanchenko}},
  \bibnamefont{and} \bibinfo{author}{\bibfnamefont{O.~I.}
  \bibnamefont{Kanakov}}, \bibinfo{journal}{Phys. Rev. E}
  \textbf{\bibinfo{volume}{73}}, \bibinfo{pages}{036618}
  (\bibinfo{year}{2006}).

\bibitem[{\citenamefont{Flach and Ponno}(2008)}]{flapon2008}
\bibinfo{author}{\bibfnamefont{S.}~\bibnamefont{Flach}} \bibnamefont{and}
  \bibinfo{author}{\bibfnamefont{A.}~\bibnamefont{Ponno}},
  \bibinfo{journal}{Physica D} \textbf{\bibinfo{volume}{237}},
  \bibinfo{pages}{908} (\bibinfo{year}{2008}).

\bibitem[{\citenamefont{Fermi et~al.}(1955)\citenamefont{Fermi, Pasta, and
  Ulam}}]{feretal1955}
\bibinfo{author}{\bibfnamefont{E.}~\bibnamefont{Fermi}},
  \bibinfo{author}{\bibfnamefont{J.}~\bibnamefont{Pasta}}, \bibnamefont{and}
  \bibinfo{author}{\bibfnamefont{S.}~\bibnamefont{Ulam}}, \bibinfo{journal}{Los
  Alamos report No LA-1940}  (\bibinfo{year}{1955}).

\bibitem[{\citenamefont{Flach and Gorbach}(2008)}]{flagor2008}
\bibinfo{author}{\bibfnamefont{S.}~\bibnamefont{Flach}} \bibnamefont{and}
  \bibinfo{author}{\bibfnamefont{A.}~\bibnamefont{Gorbach}},
  \bibinfo{journal}{Physics Reports} \textbf{\bibinfo{volume}{467}},
  \bibinfo{pages}{1} (\bibinfo{year}{2008}).

\bibitem[{\citenamefont{Galgani and Scotti}(1972)}]{galsco1972}
\bibinfo{author}{\bibfnamefont{L.}~\bibnamefont{Galgani}} \bibnamefont{and}
  \bibinfo{author}{\bibfnamefont{A.}~\bibnamefont{Scotti}},
  \bibinfo{journal}{Phys. Rev. Lett.} \textbf{\bibinfo{volume}{28}},
  \bibinfo{pages}{1173} (\bibinfo{year}{1972}).

\bibitem[{\citenamefont{Fucito et~al.}(1982)\citenamefont{Fucito, Marchesoni,
  Marinari, Parisi, Politi, Ruffo, and Vulpiani}}]{fucetal1982}
\bibinfo{author}{\bibfnamefont{F.}~\bibnamefont{Fucito}},
  \bibinfo{author}{\bibfnamefont{F.}~\bibnamefont{Marchesoni}},
  \bibinfo{author}{\bibfnamefont{E.}~\bibnamefont{Marinari}},
  \bibinfo{author}{\bibfnamefont{G.}~\bibnamefont{Parisi}},
  \bibinfo{author}{\bibfnamefont{L.}~\bibnamefont{Politi}},
  \bibinfo{author}{\bibfnamefont{S.}~\bibnamefont{Ruffo}}, \bibnamefont{and}
  \bibinfo{author}{\bibfnamefont{A.}~\bibnamefont{Vulpiani}},
  \bibinfo{journal}{J. Physique} \textbf{\bibinfo{volume}{43}},
  \bibinfo{pages}{707} (\bibinfo{year}{1982}).

\bibitem[{\citenamefont{Livi et~al.}(1985)\citenamefont{Livi, Pettini, Ruffo,
  and Vulpiani}}]{livetal1985}
\bibinfo{author}{\bibfnamefont{R.}~\bibnamefont{Livi}},
  \bibinfo{author}{\bibfnamefont{M.}~\bibnamefont{Pettini}},
  \bibinfo{author}{\bibfnamefont{S.}~\bibnamefont{Ruffo}}, \bibnamefont{and}
  \bibinfo{author}{\bibfnamefont{A.}~\bibnamefont{Vulpiani}},
  \bibinfo{journal}{Phys. Rev. A} \textbf{\bibinfo{volume}{31}},
  \bibinfo{pages}{2740} (\bibinfo{year}{1985}).

\bibitem[{\citenamefont{Luca et~al.}(1995)\citenamefont{Luca, Lichtenberg, and
  Lieberman}}]{deletal1995}
\bibinfo{author}{\bibfnamefont{J.~D.} \bibnamefont{Luca}},
  \bibinfo{author}{\bibfnamefont{A.~J.} \bibnamefont{Lichtenberg}},
  \bibnamefont{and} \bibinfo{author}{\bibfnamefont{M.~A.}
  \bibnamefont{Lieberman}}, \bibinfo{journal}{Chaos}
  \textbf{\bibinfo{volume}{5}}, \bibinfo{pages}{283} (\bibinfo{year}{1995}).

\bibitem[{\citenamefont{Berchialla et~al.}(2004)\citenamefont{Berchialla,
  Galgani, and Giorgilli}}]{beretal2004}
\bibinfo{author}{\bibfnamefont{L.}~\bibnamefont{Berchialla}},
  \bibinfo{author}{\bibfnamefont{L.}~\bibnamefont{Galgani}}, \bibnamefont{and}
  \bibinfo{author}{\bibfnamefont{A.}~\bibnamefont{Giorgilli}},
  \bibinfo{journal}{Phys. Lett. A} \textbf{\bibinfo{volume}{321}},
  \bibinfo{pages}{167} (\bibinfo{year}{2004}).

\bibitem[{\citenamefont{Luca et~al.}(1999)\citenamefont{Luca, Lichtenberg, and
  Ruffo}}]{deletal1999}
\bibinfo{author}{\bibfnamefont{J.~D.} \bibnamefont{Luca}},
  \bibinfo{author}{\bibfnamefont{A.~J.} \bibnamefont{Lichtenberg}},
  \bibnamefont{and} \bibinfo{author}{\bibfnamefont{S.}~\bibnamefont{Ruffo}},
  \bibinfo{journal}{Phys. Rev. E} \textbf{\bibinfo{volume}{60}},
  \bibinfo{pages}{3781} (\bibinfo{year}{1999}).

\bibitem[{\citenamefont{Lichtenberg et~al.}(2008)\citenamefont{Lichtenberg,
  Livi, Pettini, and Ruffo}}]{lichetal2008}
\bibinfo{author}{\bibfnamefont{A.}~\bibnamefont{Lichtenberg}},
  \bibinfo{author}{\bibfnamefont{R.}~\bibnamefont{Livi}},
  \bibinfo{author}{\bibfnamefont{M.}~\bibnamefont{Pettini}}, \bibnamefont{and}
  \bibinfo{author}{\bibfnamefont{S.}~\bibnamefont{Ruffo}},
  \bibinfo{journal}{Lect. Notes Phys.} \textbf{\bibinfo{volume}{728}},
  \bibinfo{pages}{21} (\bibinfo{year}{2008}).

\bibitem[{\citenamefont{Berchialla et~al.}(2005)\citenamefont{Berchialla,
  Galgani, and Giorgilli}}]{beretal2005}
\bibinfo{author}{\bibfnamefont{L.}~\bibnamefont{Berchialla}},
  \bibinfo{author}{\bibfnamefont{L.}~\bibnamefont{Galgani}}, \bibnamefont{and}
  \bibinfo{author}{\bibfnamefont{A.}~\bibnamefont{Giorgilli}},
  \bibinfo{journal}{Discr. Cont. Dyn. Sys. A} \textbf{\bibinfo{volume}{11}},
  \bibinfo{pages}{855} (\bibinfo{year}{2005}).

\bibitem[{\citenamefont{Benettin et~al.}(2008)\citenamefont{Benettin, Carati,
  Galgani, and Giorgilli}}]{benetal2008}
\bibinfo{author}{\bibfnamefont{G.}~\bibnamefont{Benettin}},
  \bibinfo{author}{\bibfnamefont{A.}~\bibnamefont{Carati}},
  \bibinfo{author}{\bibfnamefont{L.}~\bibnamefont{Galgani}}, \bibnamefont{and}
  \bibinfo{author}{\bibfnamefont{A.}~\bibnamefont{Giorgilli}},
  \bibinfo{journal}{Lect. Notes Phys.} \textbf{\bibinfo{volume}{728}},
  \bibinfo{pages}{151} (\bibinfo{year}{2008}).

\bibitem[{\citenamefont{Ponno and Bambusi}(2005)}]{ponbam2005}
\bibinfo{author}{\bibfnamefont{A.}~\bibnamefont{Ponno}} \bibnamefont{and}
  \bibinfo{author}{\bibfnamefont{D.}~\bibnamefont{Bambusi}},
  \bibinfo{journal}{Chaos} \textbf{\bibinfo{volume}{15}},
  \bibinfo{pages}{015107} (\bibinfo{year}{2005}).

\bibitem[{\citenamefont{Bambusi and Ponno}(2006)}]{bampon2006}
\bibinfo{author}{\bibfnamefont{D.}~\bibnamefont{Bambusi}} \bibnamefont{and}
  \bibinfo{author}{\bibfnamefont{A.}~\bibnamefont{Ponno}},
  \bibinfo{journal}{Comm. Math. Phys.} \textbf{\bibinfo{volume}{264}},
  \bibinfo{pages}{539} (\bibinfo{year}{2006}).

\bibitem[{\citenamefont{Giorgilli and Muraro}(2006)}]{giomur2006}
\bibinfo{author}{\bibfnamefont{A.}~\bibnamefont{Giorgilli}} \bibnamefont{and}
  \bibinfo{author}{\bibfnamefont{D.}~\bibnamefont{Muraro}},
  \bibinfo{journal}{Boll. Unione Mat. Ital. B} \textbf{\bibinfo{volume}{9}},
  \bibinfo{pages}{1} (\bibinfo{year}{2006}).

\bibitem[{\citenamefont{Nekhoroshev}(1977)}]{nek1977}
\bibinfo{author}{\bibfnamefont{N.}~\bibnamefont{Nekhoroshev}},
  \bibinfo{journal}{Russ. Math. Surv.} \textbf{\bibinfo{volume}{32}},
  \bibinfo{pages}{1} (\bibinfo{year}{1977}).

\bibitem[{\citenamefont{Benettin et~al.}(1985)\citenamefont{Benettin, Galgani,
  and Giorgilli}}]{benetal1985}
\bibinfo{author}{\bibfnamefont{G.}~\bibnamefont{Benettin}},
  \bibinfo{author}{\bibfnamefont{L.}~\bibnamefont{Galgani}}, \bibnamefont{and}
  \bibinfo{author}{\bibfnamefont{A.}~\bibnamefont{Giorgilli}},
  \bibinfo{journal}{Celest. Mech.} \textbf{\bibinfo{volume}{37}},
  \bibinfo{pages}{1} (\bibinfo{year}{1985}).

\bibitem[{\citenamefont{Giorgilli}(1988)}]{gio1988}
\bibinfo{author}{\bibfnamefont{A.}~\bibnamefont{Giorgilli}},
  \bibinfo{journal}{Ann. Inst. H. Poincar\'{e}} \textbf{\bibinfo{volume}{48}},
  \bibinfo{pages}{423} (\bibinfo{year}{1988}).

\bibitem[{\citenamefont{Fass\'{o} et~al.}(1998)\citenamefont{Fass\'{o}, Guzzo,
  and Benettin}}]{fasetal1998}
\bibinfo{author}{\bibfnamefont{F.}~\bibnamefont{Fass\'{o}}},
  \bibinfo{author}{\bibfnamefont{M.}~\bibnamefont{Guzzo}}, \bibnamefont{and}
  \bibinfo{author}{\bibfnamefont{G.}~\bibnamefont{Benettin}},
  \bibinfo{journal}{Commun. Math. Phys.} \textbf{\bibinfo{volume}{197}},
  \bibinfo{pages}{347} (\bibinfo{year}{1998}).

\bibitem[{\citenamefont{Guzzo et~al.}(1998)\citenamefont{Guzzo, Fass\'{o}, and
  Benettin}}]{guzetal1998}
\bibinfo{author}{\bibfnamefont{M.}~\bibnamefont{Guzzo}},
  \bibinfo{author}{\bibfnamefont{F.}~\bibnamefont{Fass\'{o}}},
  \bibnamefont{and} \bibinfo{author}{\bibfnamefont{G.}~\bibnamefont{Benettin}},
  \bibinfo{journal}{Math. Phys. Electron. J.} \textbf{\bibinfo{volume}{4}},
  \bibinfo{pages}{1} (\bibinfo{year}{1998}).

\bibitem[{\citenamefont{Niederman}(1998)}]{nie1998}
\bibinfo{author}{\bibfnamefont{L.}~\bibnamefont{Niederman}},
  \bibinfo{journal}{Nonlinearity} \textbf{\bibinfo{volume}{11}},
  \bibinfo{pages}{1465} (\bibinfo{year}{1998}).

\bibitem[{\citenamefont{Efthymiopoulos
  et~al.}(2004)\citenamefont{Efthymiopoulos, Giorgilli, and
  Contopoulos}}]{eftetal2004}
\bibinfo{author}{\bibfnamefont{C.}~\bibnamefont{Efthymiopoulos}},
  \bibinfo{author}{\bibfnamefont{A.}~\bibnamefont{Giorgilli}},
  \bibnamefont{and}
  \bibinfo{author}{\bibfnamefont{G.}~\bibnamefont{Contopoulos}},
  \bibinfo{journal}{J. Phys. A: Math. Gen.} \textbf{\bibinfo{volume}{37}},
  \bibinfo{pages}{45} (\bibinfo{year}{2004}).

\bibitem[{\citenamefont{Giorgilli}(1992)}]{gio1992}
\bibinfo{author}{\bibfnamefont{A.}~\bibnamefont{Giorgilli}},
  \emph{\bibinfo{title}{in K. Meyer, D. Schmidt, I.Cincinati and S. Dumas
  (Eds), `Hamiltonian Dynamical systems'}} (\bibinfo{publisher}{Springer},
  \bibinfo{address}{Berlin}, \bibinfo{year}{1992}).

\bibitem[{\citenamefont{Benettin et~al.}(1987)\citenamefont{Benettin, Galgani,
  and Giorgilli}}]{benetal1987}
\bibinfo{author}{\bibfnamefont{G.}~\bibnamefont{Benettin}},
  \bibinfo{author}{\bibfnamefont{L.}~\bibnamefont{Galgani}}, \bibnamefont{and}
  \bibinfo{author}{\bibfnamefont{A.}~\bibnamefont{Giorgilli}},
  \bibinfo{journal}{Phys. Lett. A} \textbf{\bibinfo{volume}{120}},
  \bibinfo{pages}{23} (\bibinfo{year}{1987}).

\bibitem[{\citenamefont{Skokos et~al.}(2007)\citenamefont{Skokos, Bountis, and
  Antonopoulos}}]{skoetal2007}
\bibinfo{author}{\bibfnamefont{C.}~\bibnamefont{Skokos}},
  \bibinfo{author}{\bibfnamefont{T.}~\bibnamefont{Bountis}}, \bibnamefont{and}
  \bibinfo{author}{\bibfnamefont{C.}~\bibnamefont{Antonopoulos}},
  \bibinfo{journal}{Physica D} \textbf{\bibinfo{volume}{231}},
  \bibinfo{pages}{30} (\bibinfo{year}{2007}).

\bibitem[{\citenamefont{Hemmer}(1959)}]{hem1959}
\bibinfo{author}{\bibfnamefont{P.}~\bibnamefont{Hemmer}},
  \bibinfo{journal}{Dynamic and stochastic type of motion by the linear chain.
  Det Physiske Seminar i Trondheim} \textbf{\bibinfo{volume}{2}},
  \bibinfo{pages}{66} (\bibinfo{year}{1959}).

\bibitem[{\citenamefont{Rink}(2001)}]{rink2001}
\bibinfo{author}{\bibfnamefont{B.}~\bibnamefont{Rink}}, \bibinfo{journal}{Comm.
  Math. Phys.} \textbf{\bibinfo{volume}{218}}, \bibinfo{pages}{665}
  (\bibinfo{year}{2001}).

\bibitem[{\citenamefont{Giorgilli}(1998)}]{gio1998}
\bibinfo{author}{\bibfnamefont{A.}~\bibnamefont{Giorgilli}},
  \bibinfo{journal}{Planet. Spa. Sci.} \textbf{\bibinfo{volume}{46}},
  \bibinfo{pages}{1441} (\bibinfo{year}{1998}).

\bibitem[{\citenamefont{Eliasson}(1997)}]{eli1997}
\bibinfo{author}{\bibfnamefont{L.}~\bibnamefont{Eliasson}},
  \bibinfo{journal}{Math. Phys. Electron. J.} \textbf{\bibinfo{volume}{2}},
  \bibinfo{pages}{1} (\bibinfo{year}{1997}).

\bibitem[{\citenamefont{Gallavotti}(1994{\natexlab{a}})}]{gal1994a}
\bibinfo{author}{\bibfnamefont{G.}~\bibnamefont{Gallavotti}},
  \bibinfo{journal}{Comm. Math. Phys.} \textbf{\bibinfo{volume}{164}},
  \bibinfo{pages}{145} (\bibinfo{year}{1994}{\natexlab{a}}).

\bibitem[{\citenamefont{Gallavotti}(1994{\natexlab{b}})}]{gal1994b}
\bibinfo{author}{\bibfnamefont{G.}~\bibnamefont{Gallavotti}},
  \bibinfo{journal}{Rev. Math. Phys.} \textbf{\bibinfo{volume}{6}},
  \bibinfo{pages}{343} (\bibinfo{year}{1994}{\natexlab{b}}).

\bibitem[{\citenamefont{Giorgilli}(1999)}]{gio1999}
\bibinfo{author}{\bibfnamefont{A.}~\bibnamefont{Giorgilli}},
  \emph{\bibinfo{title}{in C. Simo (Ed.): Hamiltonian Systems of Three or More
  Degrees of Freedom}} (\bibinfo{publisher}{Kluwer},
  \bibinfo{address}{Dordrecht}, \bibinfo{year}{1999}).

\bibitem[{\citenamefont{Giorgilli and Locatelli}(1997)}]{gioloc1997}
\bibinfo{author}{\bibfnamefont{A.}~\bibnamefont{Giorgilli}} \bibnamefont{and}
  \bibinfo{author}{\bibfnamefont{U.}~\bibnamefont{Locatelli}},
  \bibinfo{journal}{ZAMP} \textbf{\bibinfo{volume}{48}}, \bibinfo{pages}{220}
  (\bibinfo{year}{1997}).

\bibitem[{\citenamefont{R.L. et~al.}(1973)\citenamefont{R.L., Metropolis, and
  Pasta}}]{bivetal1973}
\bibinfo{author}{\bibfnamefont{R.~B.} \bibnamefont{R.L.}},
  \bibinfo{author}{\bibfnamefont{N.}~\bibnamefont{Metropolis}},
  \bibnamefont{and} \bibinfo{author}{\bibfnamefont{J.}~\bibnamefont{Pasta}},
  \bibinfo{journal}{J. Comp. Phys.} \textbf{\bibinfo{volume}{12}},
  \bibinfo{pages}{65} (\bibinfo{year}{1973}).

\bibitem[{\citenamefont{Antonopoulos and Bountis}(2006)}]{antbou2006}
\bibinfo{author}{\bibfnamefont{C.}~\bibnamefont{Antonopoulos}}
  \bibnamefont{and} \bibinfo{author}{\bibfnamefont{T.}~\bibnamefont{Bountis}},
  \bibinfo{journal}{Phys Rev E} \textbf{\bibinfo{volume}{73}},
  \bibinfo{pages}{6206} (\bibinfo{year}{2006}).

\bibitem[{\citenamefont{Flach et~al.}(2007)\citenamefont{Flach, Kanakov,
  Ivanchenko, and Mishagin}}]{flaetal2007}
\bibinfo{author}{\bibfnamefont{S.}~\bibnamefont{Flach}},
  \bibinfo{author}{\bibfnamefont{O.}~\bibnamefont{Kanakov}},
  \bibinfo{author}{\bibfnamefont{M.}~\bibnamefont{Ivanchenko}},
  \bibnamefont{and} \bibinfo{author}{\bibfnamefont{K.}~\bibnamefont{Mishagin}},
  \bibinfo{journal}{Int. J. Mod. Phys. B} \textbf{\bibinfo{volume}{21}},
  \bibinfo{pages}{3925} (\bibinfo{year}{2007}).

\bibitem[{\citenamefont{Kanakov et~al.}(2007)\citenamefont{Kanakov, Flach,
  Ivanchenko, and Mishagin}}]{kanetal2007}
\bibinfo{author}{\bibfnamefont{O.}~\bibnamefont{Kanakov}},
  \bibinfo{author}{\bibfnamefont{S.}~\bibnamefont{Flach}},
  \bibinfo{author}{\bibfnamefont{M.}~\bibnamefont{Ivanchenko}},
  \bibnamefont{and} \bibinfo{author}{\bibfnamefont{K.}~\bibnamefont{Mishagin}},
  \bibinfo{journal}{Phys. Lett. A} \textbf{\bibinfo{volume}{365}},
  \bibinfo{pages}{416} (\bibinfo{year}{2007}).

\bibitem[{\citenamefont{E.~T}(1916)}]{whi1916}
\bibinfo{author}{\bibfnamefont{W.}~\bibnamefont{E.~T}}, \bibinfo{journal}{Proc.
  R. Soc. Edinburgh} \textbf{\bibinfo{volume}{37}}, \bibinfo{pages}{95}
  (\bibinfo{year}{1916}).

\bibitem[{\citenamefont{Cherry}(1924)}]{che1924}
\bibinfo{author}{\bibfnamefont{T.~M.} \bibnamefont{Cherry}},
  \bibinfo{journal}{Proc. Camb. Phil. Soc.} \textbf{\bibinfo{volume}{22}},
  \bibinfo{pages}{510} (\bibinfo{year}{1924}).

\bibitem[{\citenamefont{Contopoulos}(1960)}]{con1960}
\bibinfo{author}{\bibfnamefont{G.}~\bibnamefont{Contopoulos}},
  \bibinfo{journal}{Z. Astrophys.} \textbf{\bibinfo{volume}{49}},
  \bibinfo{pages}{273} (\bibinfo{year}{1960}).

\end{thebibliography}

\appendix

\section{Sequence of mode excitations}

For all $q\neq q_i$, $i=1,\ldots s$, the equations yielding the
Poincar\'{e}-Lindstedt series terms $Q_q^{(1)}$ at first order are
\begin{eqnarray}
\label{k=1} \ddot {Q}_q^{(1)}+\Omega _q^2Q_q^{(1)}=-
\sum_{l,m,n=1}^{N} \Omega _q \Omega _l \Omega _m \Omega_n
C_{qlmn}Q_l^{(0)}Q_m^{(0)}Q_n^{(0)}~~~.
\end{eqnarray}
The r.h.s. of the above equations is different from zero if
$C_{qlmn}\neq 0$. In view of the definition (\ref{ccoef}) of
$C_{qlmn}$, non-zero solutions of Eqs.(\ref{k=1}) containing no new
frequencies $\Omega_q$ are possible for the values of $q$ satisfying
either $q=q^{(1)}= \mid {\pm l}\pm m \pm n \mid =m_1$ with $l,m,n
\in \{q_{1},q_{2},...,q_{s}\}$, when $1\leq m_1 \leq N$,  or
$q^{(1)}=\mid 2(N+1)- \mid {\pm l}\pm m \pm n \mid \mid$, when
$N+2\leq m_1 \leq 2N+1$ or $2N+3 \leq m_1 \leq 3N$. Both cases are
given by Eq.(\ref{app}), for $\lambda=0$ and $\lambda=1$
respectively. Thus, the proposition holds for $k=1$. Assuming it to
be true at order $k-1$, and using Eq.(\ref{last}), one finds for the
$k$th order that solutions introducing no new frequencies are
possible for the modes satisfying either $q=q^{(k)}=\mid {\pm l}\pm
m\pm n \mid $, $1\leq q^{(k)}\leq N$, or $q=q^{(k)}=\mid
2(N+1)-m_1\mid$, $1\leq q^{(k)}\leq N$, where
\begin{eqnarray} \label{lmn}
 l &=& |{2\lambda_{n_1}(N+1)-m_{n_1}|,~~
 m_{n_1}=|q_{i_1}\pm q_{i_2}\pm ...\pm q_{j_{2n_1+1}}}|,~~
 \lambda_{n_1}=\left[{m_{n_1}+N\over 2(N+1)}\right] \nonumber \\
 m &=& |{2\lambda_{n_2}(N+1)-m_{n_2}|,~~
 m_{n_2}=|q_{j_1}\pm q_{j_2}\pm ...\pm q_{r_{2n_2+1}}}|,~~
 \lambda_{n_2}=\left[{m_{n_2}+N\over 2(N+1)}\right]  \\
 n &=& |{2\lambda_{n_3}(N+1)-m_{n_3}|,~~
 n_{n_3}=|q_{r_1}\pm q_{r_2}\pm ...\pm q_{t_{2n_3+1}}}|,~~
 \lambda_{n_3}=\left[{m_{n_3}+N\over 2(N+1)}\right] \nonumber
 \end{eqnarray}
with $j_1,...,j_{2n_1+1},r_1,...,r_{2n_2+1},t_1,...,t_{2n_3+1}\in
\{1,...,s\} $, $n_1+n_2+n_3=k-1$. Taking the last relation as well
as all possible sign combinations in the sum $\pm l\pm m\pm n$ into
account, the permitted modes at $k$th order are given by equations
of the form
\begin{equation}\label{qkall}
q^{(k)} = |2(\pm\lambda_{n_1}\pm\lambda_{n_2}\pm\lambda_{n_3}\pm
g)(N+1) +(\pm q_{i_1} \pm q_{i_2}\pm \ldots \pm q_{i_{2k+1}})|,~~
\end{equation}
provided that $1\leq q^{(k)}\leq N$, where $g=0\mbox{ or }1$, and
$i_1,i_2,\ldots,i_{2k+1}\in\{1,\ldots,s\}$. After a possible sign
reversal within $|\cdot|$, not affecting the absolute value, the
expression
$$
q^{(k)} = |2(\pm\lambda_{n_1}\pm\lambda_{n_2}\pm\lambda_{n_3}\pm g)(N+1)
+(\pm q_{i_1} \pm q_{i_2}\pm \ldots \pm q_{i_{2k+1}})|
$$
always resumes the form
\begin{equation}\label{qkall2}
q^{(k)} = |2\lambda(N+1)-m_k|
\end{equation}
where $m_k=|q_{i_1} \pm q_{i_2}\pm \ldots \pm q_{i_{2k+1}}|$ and
$\lambda$ is an integer number. However, by the second restriction
of (\ref{qkall}), namely $1\leq q^{(k)}\leq N$, one necessarily has
that $\lambda=[(m_k+N)/2(N+1)]$, which concludes the proof of the
proposition.

\section{Estimates on localization profiles}

The products
$Q_{q^{(n_1)}}^{(n_1)}Q_{q^{(n_2)}}^{(n_2)}Q_{q^{(n_3)}}^{(n_3)}$
in Eq.(\ref{last}) give rise to trigonometric terms of the form
\begin{eqnarray}\label{trigterm}
Q_{q^{(n_1)}}^{(n_1)}Q_{q^{(n_2)}}^{(n_2)}Q_{q^{(n_3)}}^{(n_3)}
\rightarrow \cos(a_1\omega_1+a_2\omega_2+\ldots+a_s\omega_s)t:\nonumber\\
a_i\in{\cal Z},~~~|a_1|+|a_2|+\ldots+|a_s|\leq 2k+1
\end{eqnarray}
on the r.h.s. of Eq.(\ref{last}). For each trigonometric term of the form
(\ref{trigterm}), and $k\geq 1$, Eq.(\ref{last}) introduces a divisor to
the solution for $Q_{q^{(k)}}^{(k)}$, namely
\begin{equation}\label{div}
 \Omega_{q^{(k)}}^2-(\sum_{j=1}^s a_j\omega_j)^2=
 (\Omega_{q^{(k)}}-\sum_{j=1}^s a_j\omega_j)
 (\Omega_{q^{(k)}}+\sum_{j=1}^s a_j\omega_j)
\end{equation}
In view of Eq.(\ref{fpures}), the smallest divisors are those
satisfying
\begin{equation}\label{rescon}
q^{(k)}=\sum_{j=1}^s a_j j
\end{equation}
For such divisors, one has
$$
\omega_{q^{(k)}}-\sum_{j=1}^s a_j\omega_j=
{\pi^3\left((q^{(k)})^3-\sum_{j=1}^s a_j j^3\right)\over 24(N+1)^3}
+O\left((\pi sk/N)^5\right)
$$
while, for the sum $\sum_{j=1}^s a_j j^3$ one has the inequality
$$
|\sum_{j=1}^s a_j j^3|\leq \sum_{j=1}^s |a_j|\cdot \sum_{j=1}^s j^3
=\sum_{j=1}^s |a_j|\cdot{s^2(s+1)^2\over 4}
$$
in view of which the estimate
$$
|\sum_{j=1}^s a_j j^3|\sim {s^4\sum_{j=1}^s |a_j|\over 4}
$$
holds. On the other hand, using condition (\ref{rescon})
as well as Eq.(\ref{qomest}) one obtains
$$
(2k+1)s\sim \sum_{j=1}^s |a_j|\cdot\sum_{j=1}^s j
\sim {s^2\over 2}\sum_{j=1}^s |a_j|
$$
Combining the last two expressions we find
$$
|\sum_{j=1}^s a_j j^3|\sim {s^3(2k+1)\over 2}
$$
in view of which (combined with (\ref{qomest})) one finally arrives at
an estimate for the size of divisors
\begin{eqnarray}\label{divest}
\left|\omega_{q^{(k)}}-\sum_{j=1}^s a_j\omega_j\right|
\left|\omega_{q^{(k)}}+\sum_{j=1}^s a_j\omega_j\right|
&\sim& {\pi^3s^3(2k+1)[(2k+1)^2-{1\over2}]
\over 24(N+1)^3}\cdot{2\pi s(2k+1)\over (N+1)}\nonumber\\
&\sim&
{\pi^4s^4(2k+1)^4\over 12(N+1)^4}
\end{eqnarray}
Returning to Eq.(\ref{last}), for fixed values of $n_1$, $n_2$
and $n_3$, there are at most $s$ triplets $(q^{(n_1)},q^{(n_2)},q^{(n_3)})$
satisfying $(q^{(n_1)},q^{(n_2)},q^{(n_3)})\in{\cal D}_{q^{(k)}}$.
Using this fact as well as the estimates (\ref{divest}) and
(\ref{qomest}), one obtains for the norms of the various
terms the estimate:
$$
||{Q}_{q^{(k)}}^{(k)}||\sim
{12(N+1)^4\over \pi^4s^4(2k+1)^4}\cdot{\pi s (2k+1)\over(N+1)}\cdot
{s^4\pi^3\over (N+1)^3} \times
$$
$$
\sum_{\mathop{n_{1,2,3}=0}\limits_ {n_1+n_2+n_3=k-1}}^{k-1}
(2n_1+1)(2n_2+1)(2n_3+1)||Q_{q^{(n_1)}}^{(n_1)}||
~||Q_{q^{(n_2)}}^{(n_2)}||~||Q_{q^{(n_3)}}^{(n_3)}||
$$
or
\begin{eqnarray}\label{qkest}
||{Q}_{q^{(k)}}^{(k)}||\sim {12s\over (2k+1)^3}
\sum_{\mathop{n_{1,2,3}=0}\limits_ {n_1+n_2+n_3=k-1}}^{k-1}
(2n_1+1)(2n_2+1)(2n_3+1)||Q_{q^{(n_1)}}^{(n_1)}||~||Q_{q^{(n_2)}}^{(n_2)}||~||Q_{q^{(n_3)}}^{(n_3)}||
\end{eqnarray}
Denoting by $A^{(k)}$ the average size of the oscillations of all the modes
$q^{(k)}$, Eq.(\ref{qkest}) takes the form
\begin{eqnarray}\label{ak}
A^{(k)}= {12s\over (2k+1)^3}
\sum_{\mathop{n_{1,2,3}=0}\limits_ {n_1+n_2+n_3=k-1}}^{k-1}
(2n_1+1)(2n_2+1)(2n_3+1)A^{(n_1)}A^{(n_2)}A^{(n_3)}~~.
\end{eqnarray}
By induction it now follows that
\begin{equation}\label{anfin}
A^{(n)} \simeq {(3s/2)^nA_0^{2n+1}\over 2n+1}
\end{equation}
Indeed, assuming (\ref{anfin}) to be true for the amplitudes
$A^{(n_1)}$, $A^{(n_2)}$, $A^{(n_3)}$ in Eq.(\ref{ak}), it follows
that
$$
A^{(k)}\simeq {(3s/2)^{k-1}12sA_0^{2k+1}\over (2k+1)^3}
\sum_{\mathop{n_{1,2,3}=0}\limits_ {n_1+n_2+n_3=k-1}}^{k-1}1 =
$$
$$
{(3s/2)^{k-1}12sA_0^{2k+1}\over (2k+1)^3}\cdot{k(k+1)\over 2}
\simeq {(3s/2)^{k-1}12sA_0^{2k+1}\over (2k+1)^3}\cdot{(2k+1)^2\over 8}
= {(3s/2)^kA_0^{2k+1}\over 2k+1}~~,
$$
which demonstrates the validity of Eq.(\ref{akfin}) in the text, with
the constant $C$ having the specific value $C=3/2$.

\end{document}